\documentclass[11pt,a4paper]{iopart}
\usepackage[english]{babel}

\usepackage{iopams} 

\expandafter\let\csname equation*\endcsname\relax
\expandafter\let\csname endequation*\endcsname\relax
\usepackage{amsmath}

\usepackage{amssymb}
\usepackage{color}

\usepackage{ifthen} 
\usepackage{ifpdf}
\ifpdf
	\usepackage[pdftex]{graphicx}
	\usepackage[pdftex]{hyperref} 
	\hypersetup{colorlinks=true,
		pdfstartview=FitV,
		linkcolor=blue,
		citecolor=blue,
		urlcolor=blue,
	  pdfauthor={LIER ANDREJ liera@ipp.mpg.de},
	  pdfsubject={2020 A. Lier Alpha paper},
	  pdftitle={2020 A. Lier Alpha paper}
	}
	\DeclareGraphicsExtensions{.pdf,.png}
\else
	\usepackage[dvips]{graphicx}
	\usepackage{epsfig}
	\DeclareGraphicsExtensions{.eps}
	\usepackage{url}
\fi
\graphicspath{{figures/}}

\usepackage[numbers,sort&compress]{natbib} % Write [11-15] instead of [11, 12, 13, 14, 15], even with hyperref!
\usepackage{hypernat}

\usepackage[all]{hypcap}
\usepackage[titletoc,title]{appendix}

\newcommand{\unit}[1]{\ensuremath{\,\mathrm{#1}}}

\newcommand{\qmark}[1]{{``#1''}}
\newcommand{\vect}[1]{\mathbf{#1}}

\newcommand{\para}{\|}

\begin{document}

\title{Alpha particle driven Alfv\'enic instabilities in ITER post-disruption plasmas}

\author{A.~Lier$^1$, G.~Papp$^1$, Ph.~W.~Lauber$^1$, O.~Embreus$^2$, G.~J.~Wilkie$^3$ and S.~Braun$^4$}

%ORCIDS
% A. Lier: https://orcid.org/0000-0003-1977-2863
% G. Papp: https://orcid.org/0000-0003-0694-5446
% Ph. Lauber: none?
% O. Embreus: https://orcid.org/0000-0002-0175-5996
% G. J. Wilkie: https://orcid.org/0000-0002-6826-6273
% S. Braun: none?

\address{
$^1$Max Planck Institute for Plasma Physics, D-85748 Garching, Germany\\
$^2$Department of Physics, Chalmers University of Technology, SE-41296 Gothenburg, Sweden\\
$^3$Princeton Plasma Physics Laboratory, Princeton NJ 08540, USA\\
$^4$Center for Computational Engineering Science, RWTH Aachen University, D-52062  Aachen, Germany
}
%$^4$Center for Computational Engineering Science, RWTH Aachen University, Schinkelstr. 2, D-52062  Aachen, Germany

\ead{\href{mailto:liera@ipp.mpg.de}{Lier, Andrej <liera@ipp.mpg.de>}}
\vspace{10pt}
\begin{indented}
\item[]\date{}
\end{indented}

\begin{abstract}

\noindent Fusion-born alpha particles in ITER disruption simulations are investigated as a possible drive of Alfv\'enic instabilities. The ability of these waves to expel runaway electron (RE) seed particles is explored in the pursuit of a passive, inherent RE mitigation scenario.
The spatiotemporal evolution of the alpha particle distribution during the disruption is calculated using the linearized Fokker-Planck solver CODION coupled to a fluid disruption simulation. These simulations are done in the limit of no alpha particle transport during the thermal quench, which can be seen as a most pessimistic situation where there is also no RE seed transport. Under these assumptions, the radial anisotropy of the resulting alpha population provides free energy to drive Alfv\'enic modes during the quench phase of the disruption.
We use the linear gyrokinetic magnetohydrodynamic code LIGKA to calculate the Alfv\'en spectrum and find that the equilibrium is capable of sustaining a wide range of modes. The self-consistent evolution of the mode amplitudes and the alpha distribution is calculated utilizing the wave-particle interaction tool HAGIS. Intermediate mode number ($n=7-15,~22-26$) Toroidal Alfv\'en Eigenmodes (TAEs) are shown to saturate at an amplitude of up to $\delta B /B \approx 0.1$\% in the spatial regimes crucial for RE seed formation. We find that the mode amplitudes are predicted to be sufficiently large to permit the possibility of significant radial transport of runaway electrons.\\
\end{abstract}

\section{Introduction}
The subject of runaway electron (RE) mitigation is of crucial importance to the success of reactor-relevant tokamaks such as ITER~\cite{hender07mhd,lehnen2015disruptions,hollmann2015status,breizman2019physics,boozer2018pivotal}. Generation of REs is most concerning during disruptions, as the avalanche mechanism \cite{rosenbluth97theory,hesslow2019influence} is expected to convert a significant portion of the plasma current into runaway current on a large tokamak~\cite{boozer2015theory,martinsolis2017formation,vallhagen2020REs}. A multi-megaampere runaway beam has the potential to inflict significant damage to plasma-facing components \cite{reux2015runaway,matthews2016melt}. 
This paper discusses a phenomenon which could potentially act to passively mitigate runaway electron generation at ITER. 

The interaction of runaways with plasma waves has been investigated both in theory~\cite{fulop06destabilisation,pokol2008quasi,fulop09magnetic,komar2012interaction,komar13electromagnetic,pokol14quasi,aleynikov2015stability,liu2018role} and through observations in multiple tokamaks~\cite{zeng13experimental,papp14interaction,liu2018effects,lvovskiy2018role,heidbrink2018low,spong18first,lvovskiy2019observation}. Several tokamaks (such as DIII-D or TEXTOR ) have reported runaway suppression in correlation with increased wave activity in the current quench of the disruption or the plateau phase of the runaway beam.
Wave activities in a tokamak plasma can lead to a variety of instabilities with different effects on particle confinement to which runaway electrons -- due to their high velocity -- can be extremely susceptible to. 

In fusion, the umbrella term \qmark{shear Alfv\'en wave} collects an important type of transverse, electromagnetic plasma waves characterized by their low (Alfv\'enic) frequency range and propagation along the magnetic field. Frequency gaps in the continuum damping allow the existence of these Alfv\'en Eigenmodes (AEs). Those gaps can occur through an extremum in the safety-profile (Reversed Shear AEs~\cite{sharapov02alfven}, Global AEs~\cite{appert82GAE} and Beta-induced AEs~\cite{turnbull93BAE}) or through geometric coupling of two poloidal harmonics (Toridicity-induced AEs~\cite{cheng85TAE,cheng86TAE,heidbrink08basic}, Ellipticity-induced AEs~\cite{betti92EAE}). Compressional AEs (CAEs,~\cite{cheng01CAE,fredrickson03CAE}) are high frequency kinetic instabilities with both a perpendicular and a parallel component. 

Instabilities in the Alfv\'enic frequency range are routinely driven in fusion plasmas by energetic ions. However, in the case of such modes observed during the quench or post-disruption, the source of the drive is less obvious. Former analytical work on runaway ions~\cite{fulop14alfvenic} was inconclusive and it was later deduced~\cite{embreus15numerical} that ion runaway formation even in reactor-sized tokamaks is unlikely at disruption time scales. Spontaneous ion acceleration caused by internal magnetic reconnection events in the MAST tokamak were observed~\cite{helander02ionanalytical}. Nevertheless, the experimental observation of these modes necessitates a drive to exist. A recent publication~\cite{liu2020compressional} identified runaway electrons as a possible drive for CAEs (or possibly GAEs) on the DIII-D tokamak in the context of a Massive Gas Injection (MGI) triggered disruption and the consecutive suppression of a runaway plateau formation. 

For the ITER 15~MA scenario it is predicted~\cite{pinches15ions, schneller16alfven} that marginally unstable modes can already be present in the quiescent burning phase. Decisive for the presence of an instability however is not only the mode drive but also the strength of the competing damping. Energetic beam experiments at the AUG tokamak revealed the importance of the heavily temperature dependent Landau damping and its role in allowing strongly unstable modes to exist in a cold plasma \cite{Vannini20nled,lauber18IAEA,Horvath16nled}. 

In this paper we consider the active phase of ITER, where suprathermal alpha particles born through the fusion process in the burning plasma exist at significantly higher energies than present day experiments. This work aims at a scenario, where the post-thermal-quench healing of magnetic surfaces is fast enough to keep both runaway electrons and alpha particles confined. Good alpha particle confinement in the thermal quench stage is a necessary condition for the scenario described in the paper. This may not be universally true in all disruptions. However, if the breakup of magnetic surfaces is sufficiently strong for a sufficiently long time to cause significant alpha particle losses, then the losses of seed runaway electrons -- which possess larger thermal speeds -- will be even larger. Investigating the possible suppression of runaways via alpha-driven modes is interesting in scenarios where the runaways are reasonably confined~\cite{izzo11runaway,papp16runaway,papp19effect}, and if runaways are confined then alphas are likely confined as well. If there are some runaways being generated, the runaway current can provide the subsequent magnetic equilibrium to confine the alphas post-quench.

We show that the alpha particle distribution remains sufficiently energetic during the disruptions considered to drive TAEs in the current quench, where the plasma temperature (hence Landau damping) has dropped significantly. We also show the presence of these modes to cause significant transport to a runaway seed population. The main reason for energetic alphas to exist in this stage of a discharge is that the alpha suprathermal collision time is long compared to the thermal quench time scale. 

The paper is structured as follows. In section~\ref{sec:go} we model the evolution of the main plasma parameter profiles during the disruption using GO~\cite{feher11simulation,papp13effect,vallhagen2020REs}, which are also used to construct the magnetic equilibrium. In section~\ref{sec:codion} we describe the calculation of the spatiotemporal evolution of the alpha distribution function using the CODION tool~\cite{embreus15numerical}, and we show that a substantial suprathermal alpha population exists in the current quench. In section~\ref{sec:ligka} we describe the LIGKA model~\cite{LIGKA} used to identify the Alfv\'en spectrum and mode structures. As evidenced by simulations carried out by the relativistic version of HAGIS~\cite{pinches04role} introduced in section~\ref{sec:HAGIS}, the presence of the alpha population drives these modes to amplitudes of up to $\delta B/ B \approx 0.1\%$. Finally, in section~\ref{sec:runaway} we use HAGIS to determine the transport of seed runaways in the presence of these modes, and show seed transport.

\section{ITER natural disruption scenario \label{sec:go}}
In this paper we consider unmitigated ITER disruptions, as our aim is to investigate the possibility of inherent, ``natural'' runaway suppression mechanisms. Furthermore, the lack of mitigation significantly reduces the dimensionality of the parameter space to be considered when selecting a disruption scenario. Mitigated ITER disruptions~\cite{vallhagen2020REs} in a similar context are left for a future study. 

The pre-disruption scenario is that of the 15~MA inductive burning plasma ``scenario \#2'' described by Polevoi {et al.}~\cite{polevoi02iter,iter_scen2}, which has been extensively studied in the literature~\cite{pinches15ions,lauber15iter,schneller15ITER,schneller18ITER,schneider21alfven,papp12rmp}. High-current scenarios are also expected to produce the largest and most energetic populations of runaway electrons \cite{smith06disr}. The plasma background consists of a 1:1 mixture of deuterium and tritium. Main parameters of the pre-disruption scenario are shown in table~\ref{tab:parameters} and the temperature profiles in~\fref{fig:n-T}a.

\begin{table}
\caption{\label{tab:parameters}The plasma parameters of the pre-disruption ITER scenario.}
\begin{indented}
\item[]\begin{tabular}{@{}lll}
\br
Parameter name&Notation&Value\\
\mr
Major radius & $R_0$ & 6.195 m\\
Minor radius & $a$ & 2 m\\
Magnetic field on axis & $B$ & 5.26 T\\
Effective charge & $Z_\mathrm{eff}$ & 1.0\\
Normalised flux & $\psi$ & $\Psi(r)/\Psi(a)$\\
Normalised radius & $r/a$ & $r/a\simeq\sqrt{\psi} \equiv s$\\
Plasma current & $I_{\rm p}$ & 15 MA\\
Electron density & $n_e(\psi)$ & $10^{20}$ m$^{-3}$\\
Ion density & $n_i(\psi)$ & $10^{20}$ m$^{-3}$\\
%Density profile & $n_e(\psi)$ & Fermi \\
Electron temperature on axis & $T_{0,e}^{\rm pre}$ & 24.7 keV\\
Ion temperature on axis & $T_{0,i}^{\rm pre}$ & 21.2 keV\\
%Post-disruption temperature on axis & $T_0^{\rm post}$ & 10 eV\\
%Approx. temperature profile & $T_{\rm e}(\psi)$ & $T_0(1-0.98\psi)$\\
$q$ on axis & $q_0$ & 1.07\\
$q$ on edge & $q_a$ & 3.67\\
\br
\end{tabular}
\end{indented}
\end{table}

In order to study the post-disruption evolution of runaways and Alfv\'en waves, we need a disruption scenario and a magnetic equilibrium. As first step in constructing these, we use the GO-code~\cite{feher11simulation,papp13effect,vallhagen2020REs} to solve the 1D induction equation and to obtain the time evolution of the induced electric field $E$, the current density $j$ for both the Ohmic current and the different runaway generation mechanisms. In this paper the hot-tail~\cite{smith08hot}, the Dreicer~\cite{connor75relativistic} and the avalanche~\cite{rosenbluth97theory} mechanisms are considered. Since we are investigating a mechanism that would act on runaway seed particles, we are focusing on a scenario that is dominated by primary generation through the hot-tail mechanism. For this reason runaway seed sources coming from the active phase of ITER (i.e. tritium beta decay and inverse Compton scattering \cite{vallhagen2020REs,solis17REs}) are not considered, as they would contribute a comparatively small seed in the selected scenario.

While GO is capable of self-consistent simulation of the thermal collapse in disruptions, in order to reduce the dimensionality of the parameter space and the necessary run times for a parameter scan we are externally forcing a thermal quench: 
\begin{equation}
T (r,t) = T_{\rm f} + \left[ T (r, 0) - T_{\rm f} \right]\mathrm{exp} \left\{ - t_{\rm N} \right\}, \label{Tdrop.eq}
\end{equation}
where $T_{\rm f}$ is the final temperature, $t_{\rm N} = t/t_0$ where $t=0$ marks the instant of the thermal quench starting, $t_0$ is the exponential decay time and $T(r,t)$ is the time evolving temperature profile, equal for both electrons and ions. Since $t_{\rm N}$ is also a temperature axis for the background species, it becomes convenient to normalize the time this way and will be used throughout the sections to come. Within the GO framework we neglect impurities and alphas for simplicity (which would only have a secondary effect on the evolution through conductivity and by slightly modifying the effective charge in the Dreicer and avalanche sources). Since the thermal collapse is externally forced, we also set all plasma species at the same temperature, i.e. assuming fast equilibration between the electron and ion thermal distributions. 

\begin{figure}[htb!]
\begin{center}
\includegraphics[width = 0.46\linewidth]{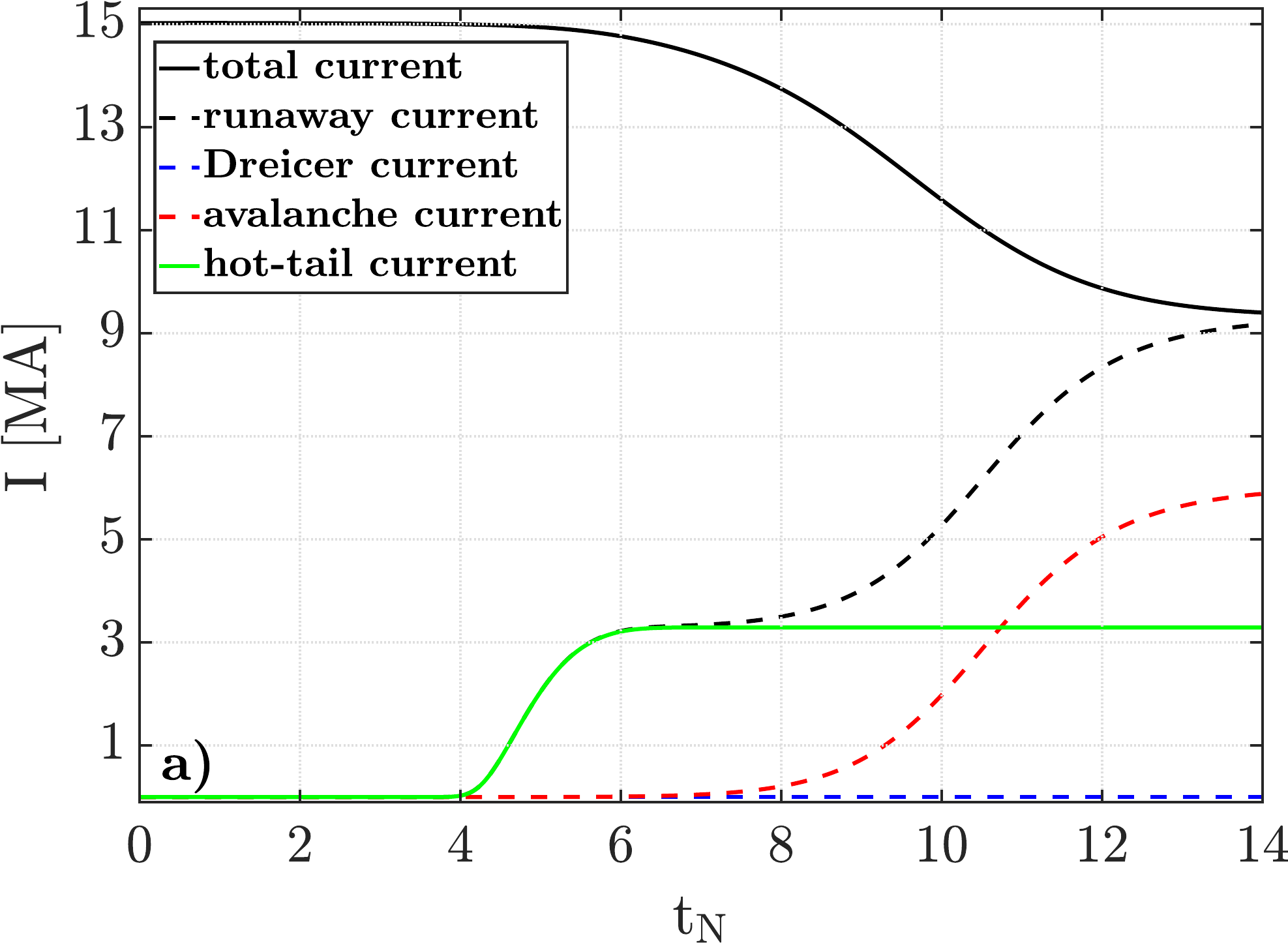} \hfill
\includegraphics[width = 0.51\linewidth]{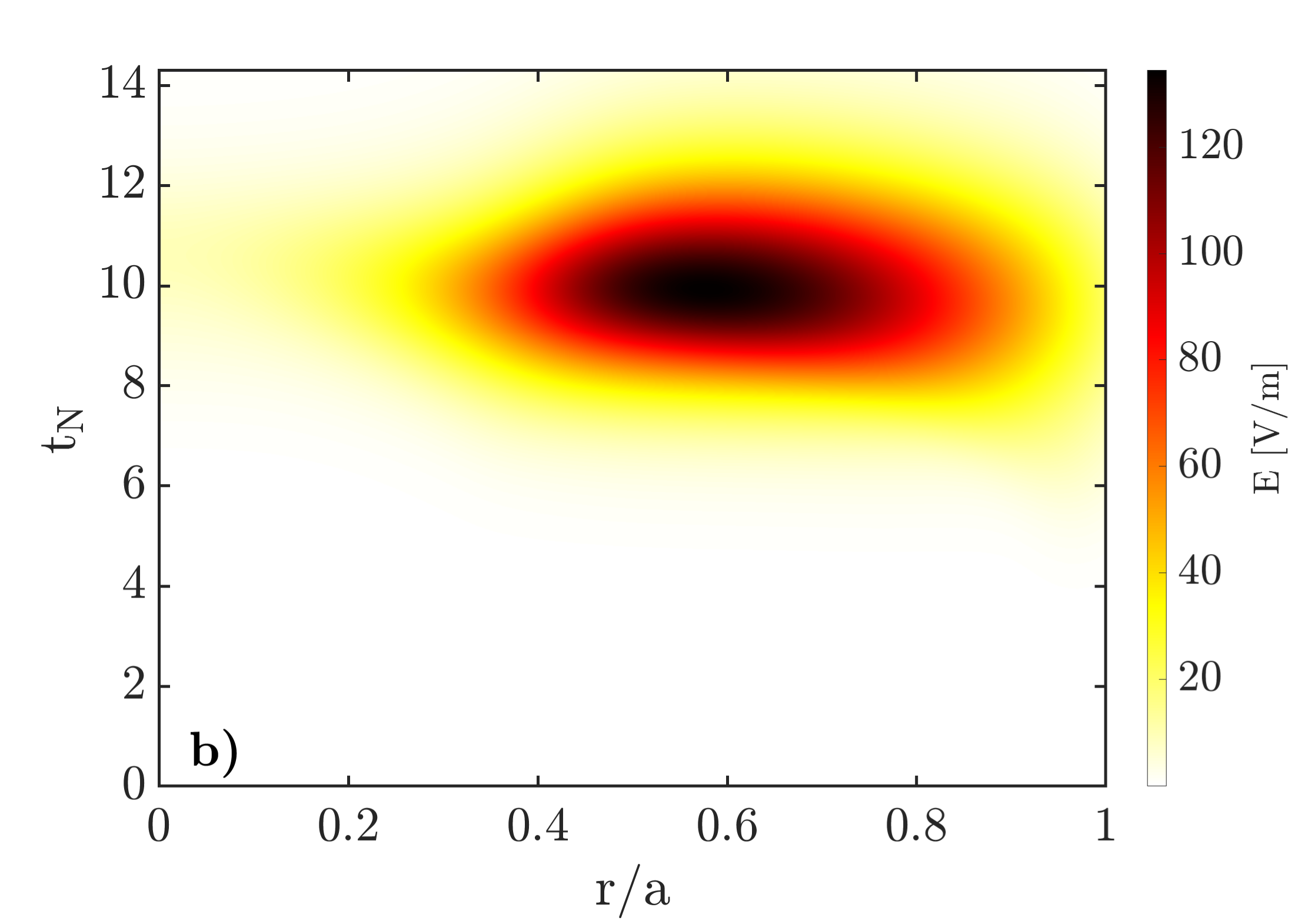}
\caption{GO-simulation of an unmitigated ITER disruption characterized by exponential temperature drop with an exponential decay time $t_0 = 0.7$~ms and a final temperature $T_f = 3$~eV. a) Plasma currents as a function of normalized time $t_{\rm N}$ categorized by the three generation mechanisms. b) The spatiotemporal evolution of the induced electric field.}
\label{fig:disr_GO}
\end{center}
\end{figure}

Figure~\ref{fig:disr_GO} depicts the GO-simulation output of a disruption identified by $T_f = 3$~eV and $t_0=0.7$~ms. A significant fraction of the current is converted to runaway through the hot-tail mechanism and due to this, little flux is available for the Dreicer mechanism. With the induced electric field setting in, the avalanche effect multiplies this initial seed. Note that with eq.~\eref{Tdrop.eq} and in a scanned range of $t_0 = 0.1-1\unit{ms}$ the hot-tail current always generates around $t_{\rm N} \approx 5$ and the electric field is induced around $t_{\rm N} \approx 8$. The induced electric field will be discussed in more detail in \sref{sec:codion} in the context of its influence on the alpha particle distribution. 

\section{Spatiotemporal evolution of the alpha particle distribution \label{sec:codion}}
In this section we are going to discuss what happens with the fusion-born alpha particle population in a disruption where a significant runaway current is still providing confinement. This question may also be important for the accurate calculation of wall heat loads during the disruption, but in this paper we are focusing on the drive to Alfv\'enic instabilities through fast ion resonances. Since mode drive can manifest both through momentum-space and real-space anisotropies, it is necessary to calculate the 1D+2V alpha particle distribution function during the disruption.

The gyro-averaged kinetic equation for the alpha particles in a homogeneous plasma with a Fokker-Planck collision operator with the source term $S_\alpha$ for the fusion of alpha particles is
\begin{equation}
\frac{\partial f_\alpha}{\partial t} + \frac{e Z_\alpha}{m_\alpha} E \left(\xi \frac{\partial}{\partial v} + \frac{1 - \xi^2}{v} \frac{\partial}{\partial \xi} \right) f_\alpha = \sum_s \mathcal{C}_{\alpha,s} \lbrace f_\alpha \rbrace + S_\alpha,
\label{reducedionkinetic.eq}
\end{equation}
where $f_\alpha$ is the alpha particle distribution function, $e$ the elementary charge, $Z_\alpha = 2$ is the alpha particle charge number, $m_\alpha$ the alpha particle mass and $\mathcal{C}_{\alpha, s}$ the linearized collision operator  describing collisions with species $s$. The particle pitch $\xi = v_\|/v$ is defined with respect to the equilibrium magnetic field lines. 

Equation~\ref{reducedionkinetic.eq} is a reduction of the ion kinetic equation, computing the time evolution of a distribution function in velocity space ($v,\xi$) under the influence of electric field acceleration, the Lorentz force and an accumulation of small-angle Coulomb collisions. Trading off spatial features such as neoclassical particle transport is justified by the charged particle motion in a tokamak being dominated by parallel dynamics because of the preferred direction of the conductivity ($\sigma_\| \gg \sigma_\perp$). Perpendicular dynamics will be discussed later on. The magnetic field strength does not enter as a quantity into the equation, as at the energies investigated Bremsstrahlung and synchrotron emission losses can be neglected for the alpha particles.

As discussed earlier, we consider a 1:1 mixture of deuterium and tritium. The fusion process births alpha particles isotropically, at an energy of $E_\alpha= 3.5~\mathrm{MeV}$. The empirically derived reaction rate for Deuterium and Tritium is (NRL~\cite{NRL})
\begin{equation*}
\langle\sigma v\rangle_{\mathrm{DT}} = 3.68 {\cdot} 10^{-12}\, T_i^{-2/3} \mathrm{exp} \left\{ -19.94~T_i^{-1/3} \right\}~\mathrm{cm}^{-3} \mathrm{s}^{-1},
\end{equation*}
where $T_i$ is the temperature of both deuterium and tritium given in keV and must be below 25~keV for the formula to be valid. The reaction rate for deuterium-deuterium fusion is not included, as it is generally two orders of magnitude lower. With equal D-T ion densities $n_{\rm D,T}$ and temperatures $T_i$, transforming from the fusion reaction center-of-mass frame to the lab frame results in the following energy dependence of the alpha particle source~\cite{brysk73fusion}:
\begin{equation}
S_\alpha  = S_0 ~ \mathrm{exp} \left\{ -\frac{5}{16} \frac{ \Big(\langle E_{\alpha}\rangle -E_{\alpha}\Big)^2}{T_i \langle E_{\alpha} \rangle} \right\}.
\label{source.eq}
\end{equation}

%Assuming equal D-T ion densities $n_{\rm D,T}$ and a Gaussian-distribution of the birth energy $\langle E_{\alpha}\rangle$, the alpha particle source in the kinetic equation equation~(\ref{reducedionkinetic.eq}) becomes
%\begin{equation}
%S_\alpha  = S_0 ~ \mathrm{exp} \left\{ -\frac{5}{16} \frac{ \Big(\langle E_{\alpha}\rangle -E_{\alpha}\Big)^2}{T_i \langle E_{\alpha} \rangle} \right\}
%\label{source.eq}
%\end{equation}
%with the source magnitude $S_0 = n_{\rm D,T}^2 \langle\sigma v\rangle_\mathrm{DT}$. The source distribution in energy has been calculated by Brysk~\cite{brysk73fusion} with the factor $5/16$ originating from the transformation of the center of motion frame of the ions to the lab frame.

We solve the kinetic equation~(\ref{reducedionkinetic.eq}) numerically with the tool CODION (COllisional Distribution of IONs)~\cite{embreus15numerical}. CODION was originally developed to study highly energetic ion runaway mechanisms~\cite{gibson59ionrunaway} in fusion and astrophysical plasmas. It can be considered an extension of previously existing analytical models~\cite{helander02ionanalytical} for the cases of low electric field and trace impurities. CODION uses a linearized collision operator, which is valid as long as the density of alpha particles is sufficiently small compared to the background.

\begin{figure}
\begin{center}
\includegraphics[width=0.515\linewidth]{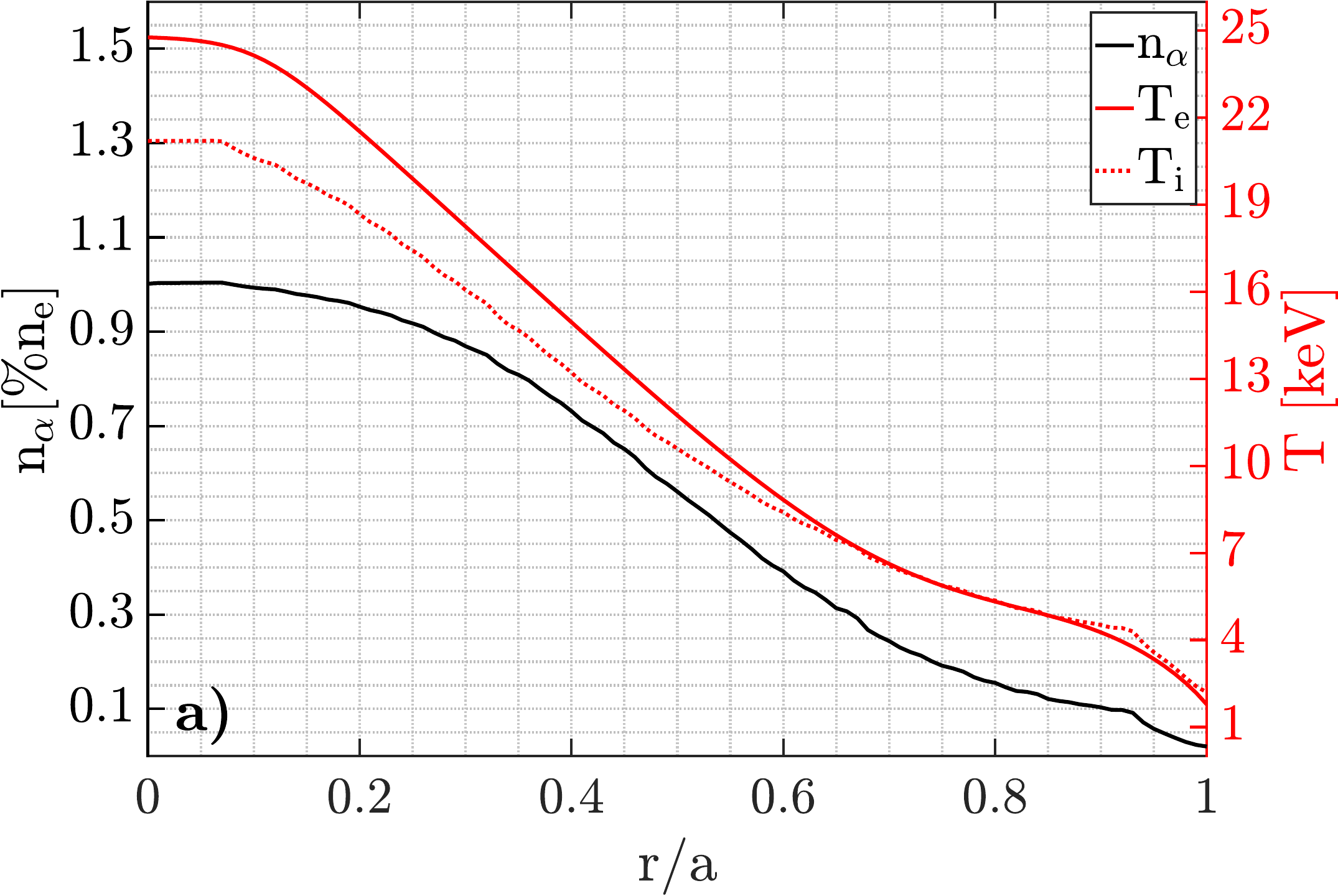} \hfill
\includegraphics[width=0.465\linewidth]{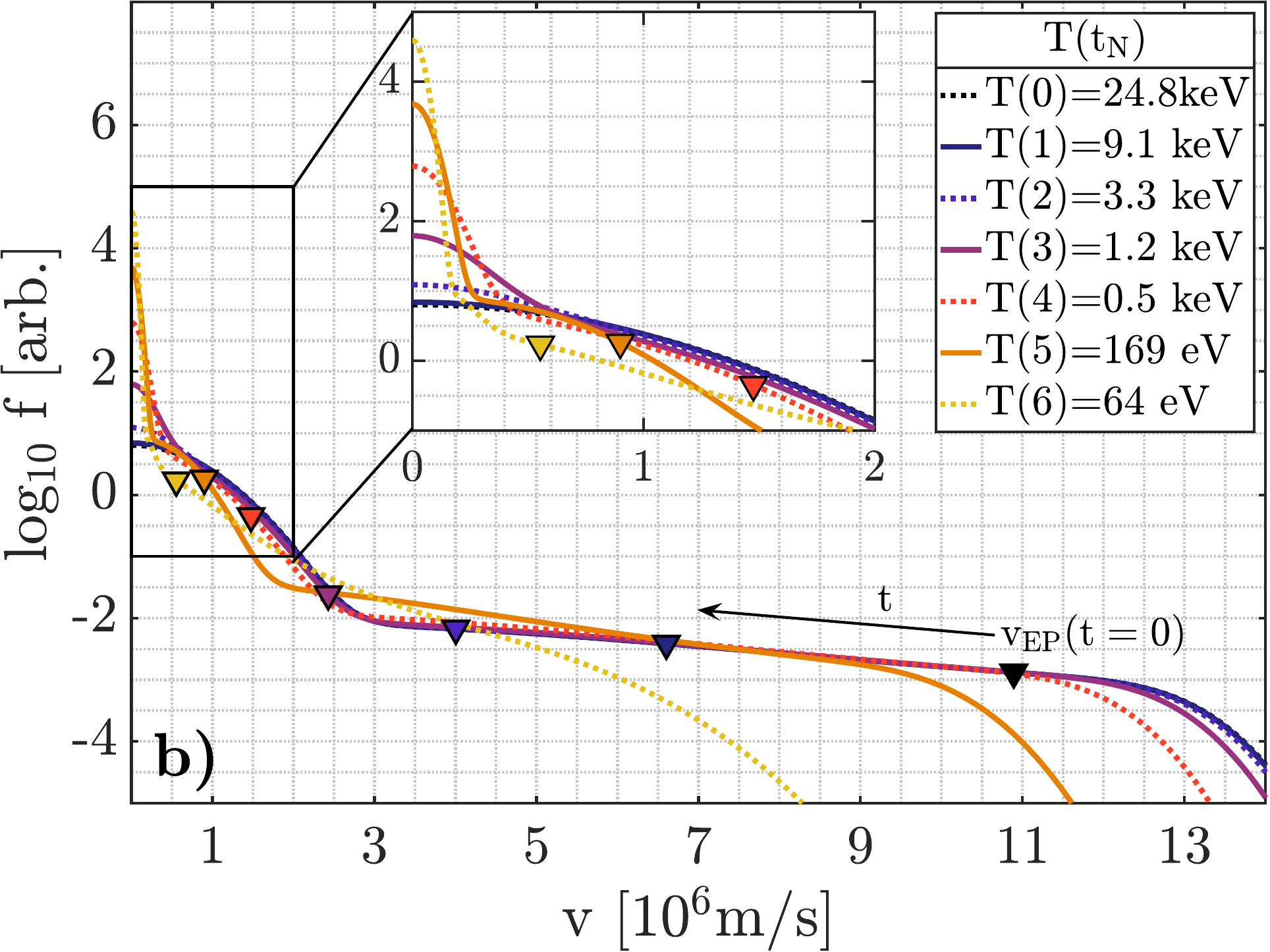}
\caption{a) Pre-disruptive ion and electron temperature profiles and the alpha particle density $n_\alpha$ established through numerical fusion in CODION. b) Time evolution of the alpha velocity distribution on-axis during a disruption defined by $T_f = 3~$eV and $t_0=0.7$~ms and as a function of $T(t_{\rm N} = t/t_0)$. The alpha particles are born at $v_\alpha \approx 13 \cdot 10^6 \unit{m/s}$ with a spread $\Delta v$ corresponding to the pre-disruptive electron background temperature $T_{0,e}^{\rm pre} = 25$~keV. The temperature dependent velocities $v_{\rm EP}$ above which particles are considered energetic are marked as a triangle of the corresponding color. Also included are the Alfv\'en velocities $v_{\rm A}$ for the plasma composition chosen. Note that the total density integral is conserved throughout the disruption as the fusion process comes to a hold and particle losses are ignored.}
\label{fig:n-T}
\end{center}
\end{figure}

In order to compute the radial distribution, independent CODION calculations are executed on a radial grid spanning the plasma radius with 101 points. The initial steady-state alpha distribution is established self-consistently by providing the initial profiles of temperature $T_{\rm s}(r)$ (see \fref{fig:n-T}a) and densities $n_{\rm s}(r)$. We take the electron density as radially flat and constant at a value of $n_{\rm e} = 10^{20}$~m$^{-3}$ and the 50:50 fuel ion densities $n_{\rm D,T}$ to fulfill quasi-neutrality. The alpha density $n_\alpha$ and pressure $p_\alpha$, which is necessary for later calculations, are evaluated (assuming isotropy and applying gyro-averaging) with the use of moments
\begin{align}
p_\alpha (r,t) & = \frac{4 \pi m_\alpha}{3}\int  v^4 f_\alpha(r,v,t) \label{p.eq} \rmd v,\\
n_\alpha (r,t) & = 4 \pi \int  v^2 f_\alpha(r,v,t) \rmd v. 
\label{n.eq}
\end{align}
Isotropy is assumed as the birth process is isotropic, and the pre-disruptive electric field is small, therefore for the initialization simulation set to zero. The integrals are computed with the use of Simpson quadrature weights. For later use, we further specify the energetic part of the alpha population by limiting the lower bound of the integrals to $v_{\rm EP} = 10\sqrt{2T(r,t)/m_\alpha}$ and thus define energetic particle density $n_{\alpha,{\rm EP}}$ and energetic particle pressure $p_{\alpha,{\rm EP}}$. This definition is not suited for pre-disruption analysis (as the cut-off velocity would be too high), however, it is purposeful for our post-disruptive analysis. Starting with initially negligible alpha particle densities $n_\alpha$ (for numerical reasons) we simulate the fusion process $S_\alpha$ (eq.~\eref{source.eq}) self-consistently until a desired $n_\alpha (r=0) = 0.01n_e$ is reached. The alpha profile established by our simulation is shown in \fref{fig:n-T}a. We do not remove a fused D/T atom from the ion density profile nor do we add the born alpha particle to it, as it remains a minority species. Targeted for ITER is an optimal 50:50 D:T mixture and the efficiency of the pump-out of He-ash is not yet clear. Radial particle transport is not captured by the code. However, since we consider situations of good confinement in which the alpha particle loss time is approximately three orders of magnitude higher than its slowing down time, this seems to be a good approximation.

In velocity space, after sufficient simulation time, the alpha particles form a slowing-down distribution~\cite{gaffey76energetic}
\begin{equation}
f_{\rm SD} (v) = \frac{C}{v_c^3 + v^3}~\mathrm{Erfc}\left( \frac{v - v_\alpha}{\Delta v} \right),
\label{f_SD.eq}
\end{equation}
where $C$ is a constant proportional to $S_\alpha$, $\mathrm{Erfc}(x)$ is the complementary error function, $v_\alpha = \sqrt{2E_\alpha/m_\alpha}$ is the birth velocity, $v_c$ is the crossover velocity and $\Delta v$ is the velocity spread of the fusion reactants at birth. In the absence of a particle sink, the alphas born at $v_\alpha$ eventually thermalize into a Maxwellian $f_M$ of background temperature (helium ash). The total alpha distribution therefore consists of a slowing-down part and a thermal Maxwellian $f_\alpha = f_{\rm M} + f_{\rm SD}$. Using this, we fit the CODION simulated time-dependent distributions and determine $v_c$, $\Delta v$ and $C$ for further analytical processing of the energetic tail ($v>v_{\rm EP}$), which we assume to be the described fully by the slowing-down part $f_{\rm SD}$. In steady state these coefficients can be determined through the given plasma parameters.

In simulating the effect of the disruption, the initial steady-state alpha particle distribution is subjected within CODION to the time varying background profile of temperature as described in \sref{sec:go}. Effects of the induced electric field will be neglected since we only calculate $f_\alpha$ up to $t_{\rm N} = 6$. Including the electric field in this duration is more expensive numerically, and only leads to a sub-1\% change in the fast particle pressure. The high-energy alphas exist during the current quench due to their relatively slow deceleration, rather than due to electric field acceleration. For the disruption the fusion source is disabled, conserving the total density and density profiles. Background populations are assumed to maintain thermal equilibrium, while the collisional cooling of the alpha particles is consistently calculated by CODION. The disrupting alpha distribution function for the chosen example case $T_f = 3~$eV and $t_0=0.7$~ms is shown in \fref{fig:n-T}b. The energetic tail of alphas is largely conserved during simulation time, due to the short background cooling time compared to the collision time of the fast alphas. The analytically derived~\cite{spitzer62book} slowing-down time for an alpha particle colliding with the steady state background on ITER is on the order of a second~\cite{fasoli07chapter} and drops to the order of milliseconds for $T \approx 1 \unit{keV}$. Additionally marked in \fref{fig:n-T}b are the velocities $v_{\rm EP}$ above which the particles are considered energetic and also the Alfv\'en velocity $v_{\rm A} = B / \sqrt{\mu_0 \rho}$, with the mass density $\rho$, the magnetic field strength $B$ and $\mu_0$ the permeability. Note how $v_{\rm EP}(t_{\rm N}=3)$ separates the Maxwellian from the energetic tail, which we assume to be fully described by $f_{\rm SD}$. 

\begin{figure}[htb!]
\begin{center}
\includegraphics[width = 0.51\linewidth]{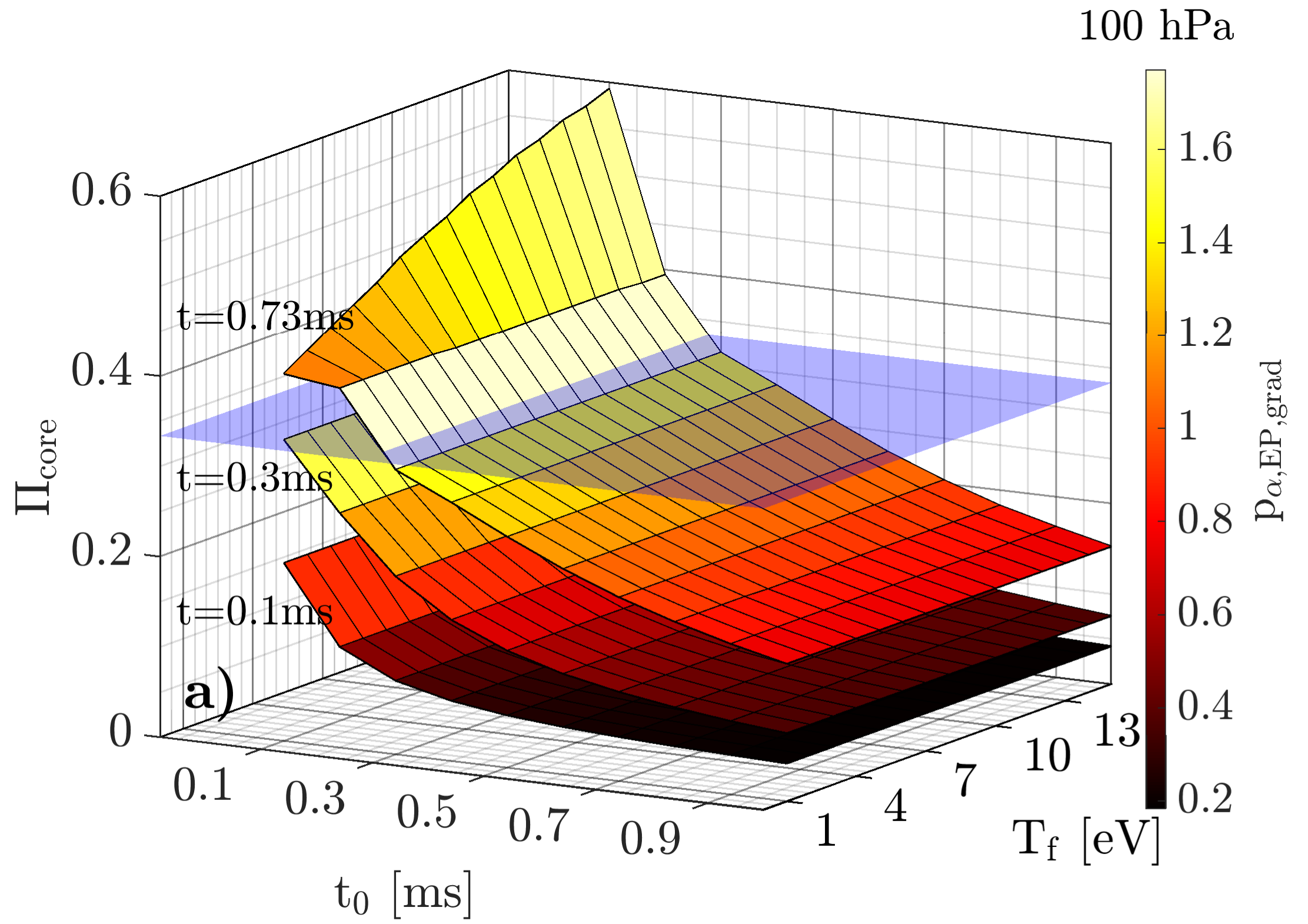} \hfill
\includegraphics[width = 0.47\linewidth]{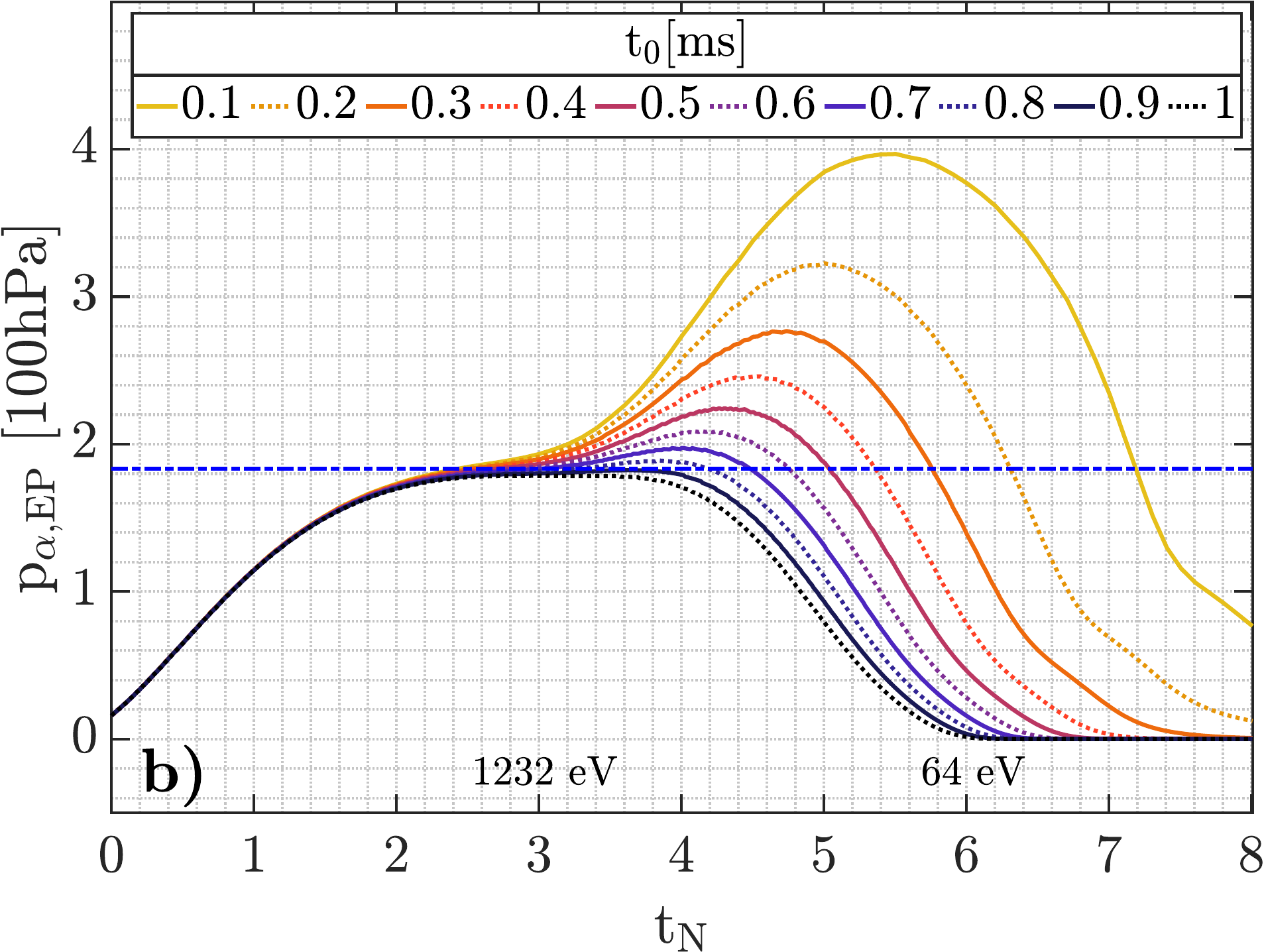}
\caption{a) Scan over the $t_0$-$T_f$ parameter space showing time-slices of the fraction of energetic to total alpha particle pressure $\Pi_{\rm core}$ in the core, as computed by the moments applied onto the CODION-simulated distribution functions. The colors show $p_{\alpha,\rm EP,\rm grad}$ and the blue plane marks the pressure level reached by the chosen disruption scenario ($T_f = 3 \unit{eV}$, $t_0 = 0.7 \unit{ms}$) after 2.1\unit{ms}. b) Time-evolution of the $T_f = 3 \unit{eV}$ slice of the scan for various $t_0$ in the core. The blue dotted line corresponds to the blue plane in the left figure. Due to the temperature quench being described by eq.~\eref{Tdrop.eq}, the x-axis can be interpreted as a background temperature.}
\label{fig:Pi_scan}
\end{center}
\end{figure}

We conduct a scan over various disruption scenarios defined by $T_f = [1-15]$~eV and $t_0 = [0.1-1]$~ms. The presence of an energetic population we quantify as the ratio of energetic to total pressure $\Pi \equiv p_{\alpha,\rm EP}/p_\alpha$ and indicate the ability to drive modes via its gradient in real space. \Fref{fig:Pi_scan}a shows the time-slices of (normalized) core pressures and uses colorcode for the energetic particle pressure gradient $p_{\alpha,\rm EP,\rm grad} = p_{\alpha,\rm EP}(r/a=0) - p_{\alpha,\rm EP}(r/a=0.6)$. The general behaviour of $\Pi_{\rm core}(t)$ and its gradient in this plot can be imagined as a wave propagating towards higher $t_0$ with the effect of lower $T_f$ forcing earlier drops in pressure, as indicated by the time-slice for $t=0.73\unit{ms}$. The peak of this \qmark{wave} gradually drops after reaching up to 80\% for the quickest thermal quench simulated (note that the denominator in $\Pi$ does not contain electron and background ion pressure). 
\Fref{fig:Pi_scan}b depicts the $t_0$ parameter space for $T_f = 3\unit{eV}$ and the energetic pressure in the core in absolute numbers. The general rise in $p_{\alpha, \rm EP}$ is due to the definition of $v_{\rm EP} \propto \sqrt{T}$ and what is considered by us to be \qmark{energetic} in reference to the background (compare to $v_{\rm EP}$ in \fref{fig:n-T}b). Its magnitude being a product of particle densities and energies serves as an indication to the remnant of the tail throughout the disruption. The interesting result is found in difference in rise ($3 < t_{\rm N} $) and the lack of difference in rise  ($0 < t_{\rm N} <  3$). First, pressure evolution shares nearly identical behavior for the scanned range and under the assumptions made (exponential temperature decay, pure D-T composition, instantly thermalizing background species). In the context of eq.~\eref{Tdrop.eq} $t_{\rm N}$ becomes a temperature axis of the background species, thus a collision time-scale. Exemplary shown in \fref{fig:n-T}b (for $t_0 = 0.7\unit{ms}$) was the barely changing alpha distribution until $t_{\rm N}=3$. The qualitatively different evolution afterwards suggests a deviation caused by a growing discrepancy of elapsed time $t=t_{\rm N} t_0$ to collisionality. As expected, the quicker and more violent thermal quenches sustain a more significant energetic alpha particle tail in reference to the background temperature. This similarity for $t<t_{\rm N}=3$ for the core pressure hold qualitatively true up to a radial point of $r/a \approx 0.5$ in our simulations. In the outer half of the plasma the background temperature to begin with is low enough to collisionally drag the energetic tail early on.
%A more exact evaluation of the parameter space is left for future work. 

In the absence of a bump-on-tail drive (due to the electric field not yet setting in) we evaluate the possibility of drive through the pressure gradient of the alphas. For a preferably general result we opt to evaluate the distributions and the pressure they exert at $t_{\rm N} =3$. Our definition of $v_{\rm EP}$, which separates $f_{\rm M}$ and $f_{\rm SD}$, makes sure that the energetic tail can optimally be described by the slowing down distribution equation for the time-point chosen. As is shown in figure \fref{fig:n-T}b important TAE resonance regions for energies above $100 \unit{keV}$ \cite{schneller13alfv} are populated at this point in time.

\begin{figure} 
\begin{center}
\includegraphics[width = 0.49\linewidth]{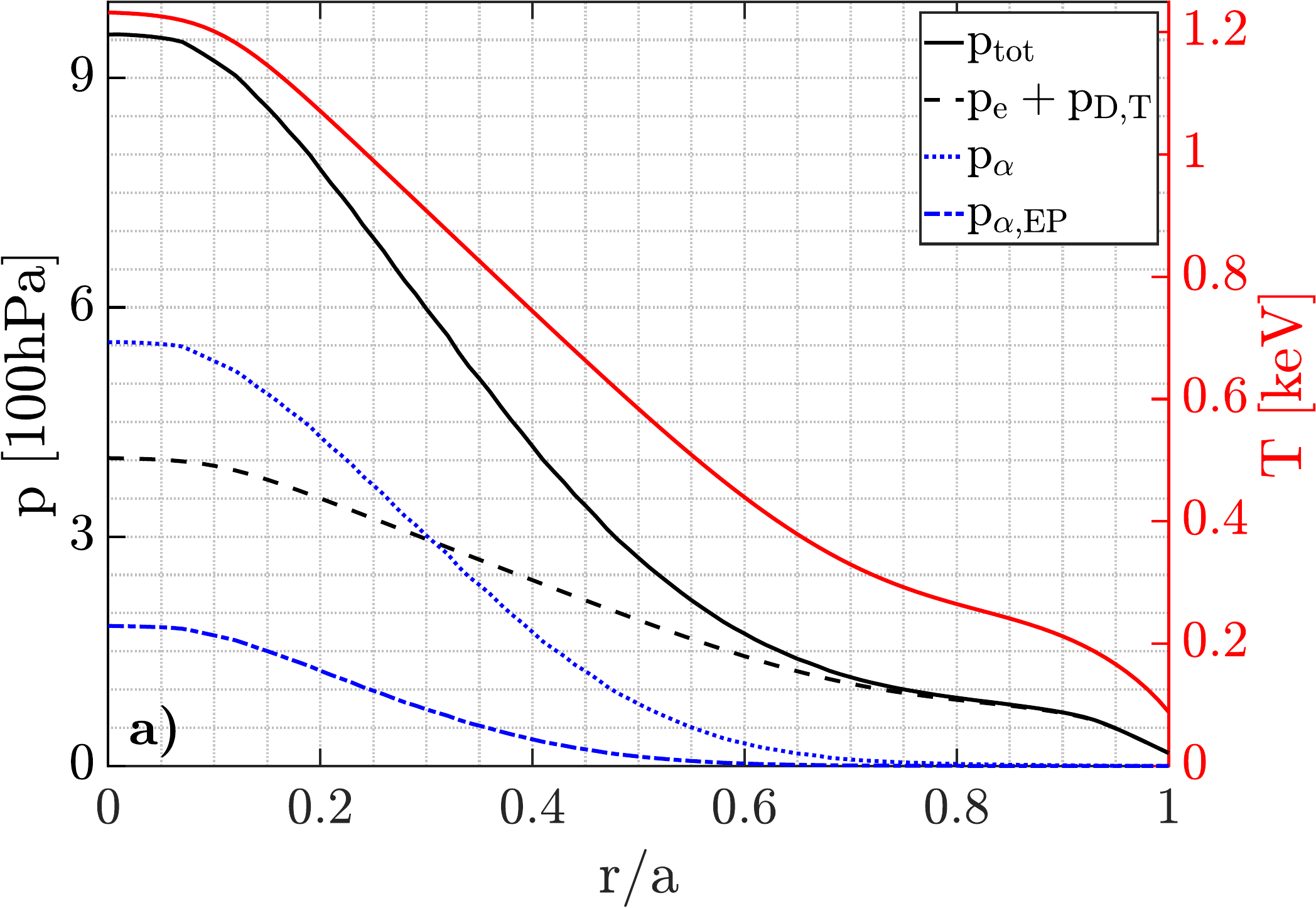} \hfill
\includegraphics[width = 0.49\linewidth]{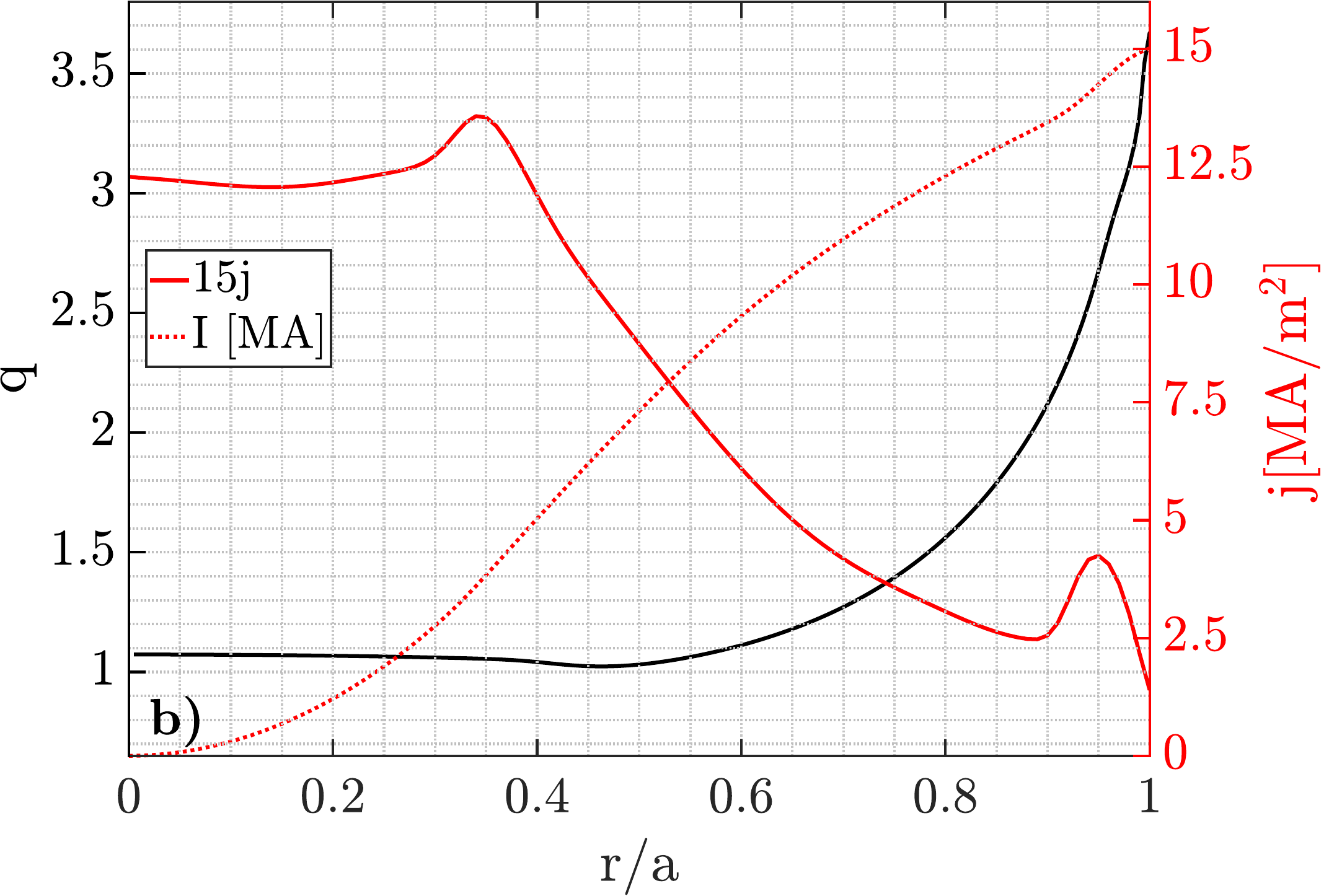}
\caption{a) Pressures and temperature profiles $2.1\unit{ms}$ into the chosen disruption ($t_0 = 0.7\unit{ms}$). Alpha pressures $p_\alpha$ and $p_{\alpha, {\rm EP}}$ are determined by velocity moments of the CODION simulated distribution functions, while the electron ($p_e$) and ion ($p_{\rm D,T}$) pressures are computed using the ideal gas law at respective temperatures, which evolved with eq.~(\ref{Tdrop.eq}). b) Current density $j$, integrated current $I$ and the safety factor profile $q$ at $t_{\rm N} = 3$ as calculated by GO. At this point in time, the current is still dominated by the Ohmic contribution as runaway electrons are yet to be generated (see \fref{fig:disr_GO}a).}
\label{fig:q_profile}
\end{center}
\end{figure}

\Fref{fig:q_profile}a shows the individual species' pressures and temperature profile at $t_{\rm N} = 3$ (for the selected case of $t_0=0.7\unit{ms}$ and $T_f=3 \unit{eV}$) and is going to be used to reconstruct the equilibrium. The total pressure is determined as $p_{\rm tot} = p_e + p_{\rm D,T} + p_\alpha$, where electron pressure $p_{\rm e}$ and deuterium/tritium pressure $p_{\rm D,T}$ are calculated from the ideal gas equation. Though the alpha particles are a minority species in the plasma, they contribute significantly to the pressure and its gradient due to their large kinetic energy. In initial steady state, the alpha pressure contributed to $\approx 10\%$ of the total pressure on axis. With the thermal background cooling rapidly its relative relevance grows. It even briefly dominates in the core before vanishing together with the energetic pressure at $ t > 6 t_{\rm N}$ (compare to \fref{fig:Pi_scan}b). Note that the Alfv\'en mode drive we are investigating in the following sections is determined by the actual distribution functions $f_{\rm SD}(r)$ and not simply by their integral moments.

\section{The Alfv\'en spectrum in the current quench\label{sec:ligka}}
In order to determine the Alfv\'en spectrum, it is necessary to reconstruct the equilibrium. The characteristic time point chosen during the disruption is set at $t_{\rm N} = 3$. On the one hand there is still a substantial population of alpha particles, on the other hand the core background temperature has fallen to $\approx 1232$~eV, which means that various mode damping mechanisms are also lowered. The plasma equilibrium is constructed using the pressure profiles (\fref{fig:q_profile}a) and the current density profiles from GO (\fref{fig:q_profile}b) using the VMEC code~\cite{VMEC}. As aforementioned, the pressure profiles are very similar for the various disruption scenarios up to this time-point and since no runaway current is generated yet (see \sref{sec:go}) the same holds true for the current density profiles. The total time-window for the mode-evolution is going to be $3t_{\rm N}< t < 6t_{\rm N}$, during which we will assume the plasma equilibrium to be constant. The low post-disruption pressure has diminishing influence on shaping the equilibrium and the current density profile is essentially constant in this time-frame, since the current quench is just about to begin (see \fref{fig:disr_GO}a).

\begin{figure}[htb]
\begin{center}
\includegraphics[width = 0.50\linewidth]{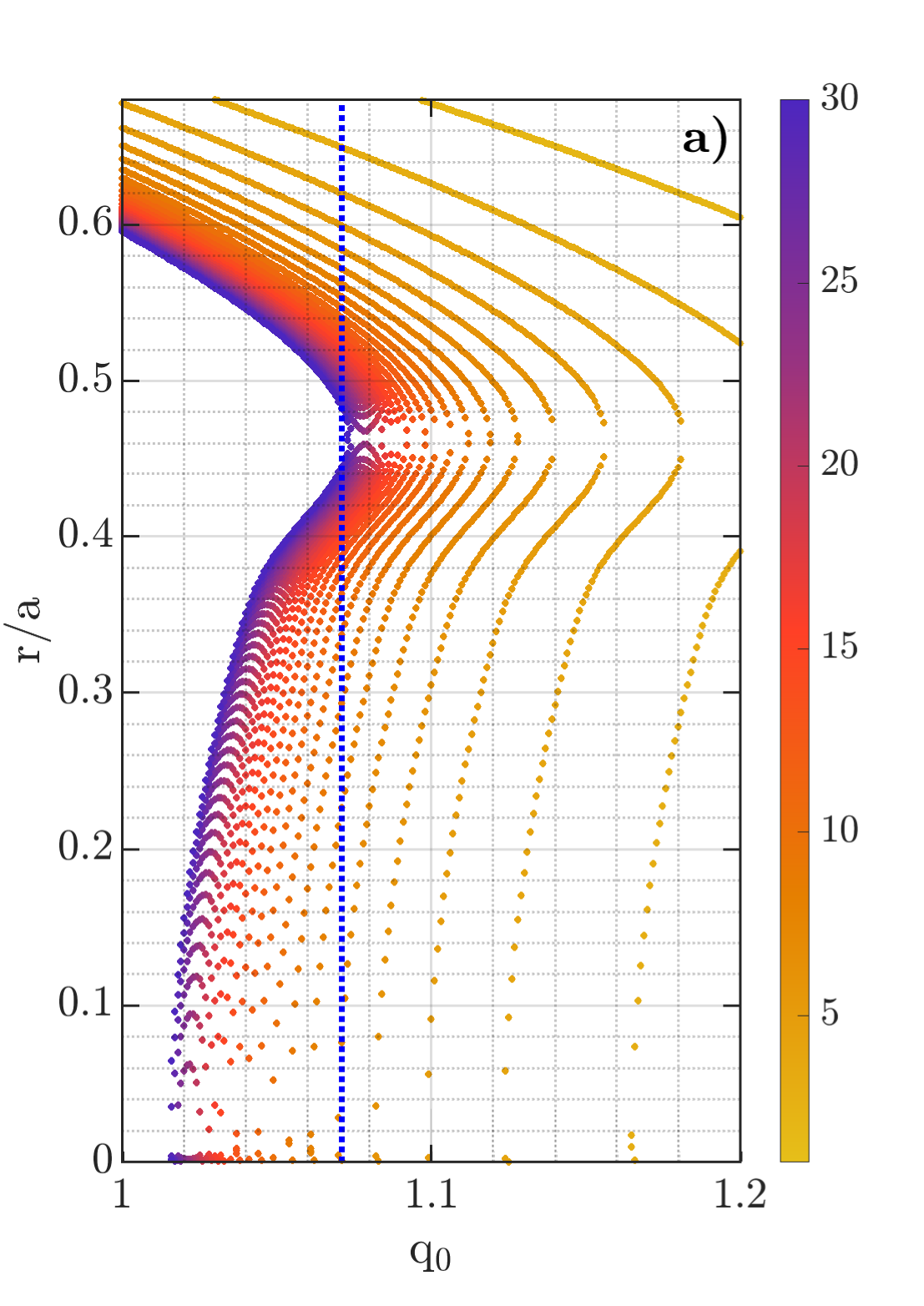} \hfill
\includegraphics[width = 0.49\linewidth]{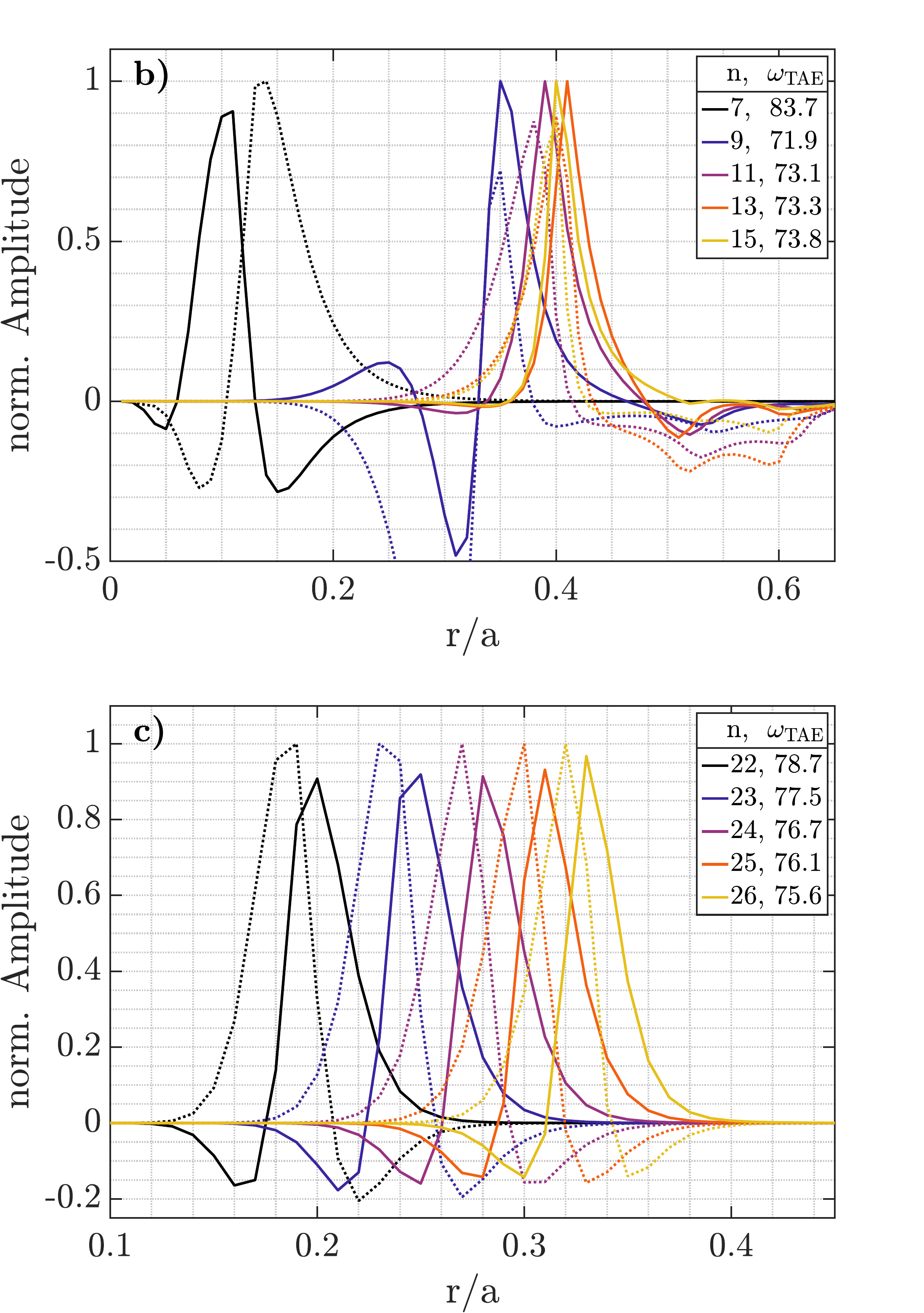}
\caption{a) Scan over $q_0$ of the possible TAEs $t_N=3$ into the chosen ITER disruption ($T_f = 3\unit{eV}$, $t_0 = 0.7\unit{ms}$) as allowed by the current density and pressure profiles. Depicted by color are TAEs of toroidal mode number $1 < n < 30$. The blue dotted line represents the $q_0$-value chosen for further analysis. RHS: Real part of the normalized mode structure of b) $n = m, m+1$ and c) $n=m+1,m+2$ coupled TAEs as computed by LIGKA for the post-disruptive ITER plasma ($t_{\rm N} = 3$). Mode frequencies $\omega_{\rm TAE} [\unit{kHz}]$ are provided in the legend. }
\label{fig:modes}
\end{center}
\end{figure}

In order to evaluate the Alfv\'en spectrum, as well as the frequency and mode structure of modes possibly supported by this equilibrium, we employ the LIGKA code (LInear GyroKinetic Alfv\'en physics)~\cite{LIGKA}. We carry out a scan for the absolute scaling of the safety factor to account for deviations in the scenario and the time evolution of the plasma current, while maintaining the shape (determined by the plasma background). \Fref{fig:modes}a shows the results of this scan and reveals frequency gaps for TAEs of toroidal mode numbers $1 < n < 30$. Due to the alpha particle orbit width in ITER the toroidal mode numbers are high compared to present day tokamaks. The solutions shown are from the even TAE branch that have been found to be the most unstable AEs in previous ITER analyses~\cite{pinches15ions}. Also, Beta-induced Alfv\'en Eigenmodes (BAEs) are found to be present in the steep pressure region, however not further considered in this work. TAEs and BAEs are often deemed dominant considering the particle transport they cause due to their generally low frequency hence higher (potential) particle displacement per wave-particle energy transfer \cite{chen88icrf}. We chose the nominal value of the core safety-factor to be $q_0=1.071$ with the toroidal harmonic ranges $n = [7-15,~22-26]$ and respective poloidal harmonics $m = \left[\{(n-2) - (n+4)\},~\{(n-2) - (n+6)\}\right]$ as justified by the core-localization. The center of the mode frequency gap is located at $\omega_{\rm TAE} = v_{\rm A}/(4 \pi q R) \propto 1/(q \sqrt{n_{\rm e}})$ and varies mainly due to the $q$-profile, since the electron profile is flat. The corresponding (normalized) radial structures are shown in \fref{fig:modes}b-c. Our modeling shows that most of the modes possibily sustained by the post-disruption equilibrium are core localized, which is beneficial for interaction with the mainly core localized runaway electron seed. The Alf\'en velocity $v_{\rm A}$ has a value of $v_{\rm A}\simeq 7.3 \cdot 10^6 \unit{m/s}$ for the plasma composition chosen. For TAEs the most fundamental resonances occur at $v=v_{\rm A}$ and $v=v_{\rm A}/3$ \cite{heidbrink08basic}, which is still populated by the energetic alpha tail in velocity space at $t_{\rm N}<6$ (\fref{fig:n-T}b).

\newpage

\section{Interaction of alpha particles and Alfv\'en waves \label{sec:HAGIS}}
The time evolution of the modes and their saturation amplitude is a critical question to determine their potency for runaway transport. Earlier studies showed that a magnetic perturbation with an amplitude of about $\delta B / B \approx 0.1\%$ is sufficient to suppress runaway avalanche~\cite{helander00suppression,papp15magper}, while more recent research~\cite{svensson20magper} decreases this threshold by about a factor of 2.

The interaction of the modes with the alpha particles (providing the drive) and the runaways is calculated using HAGIS~(HAmiltonian GuIding center System)~\cite{HAGIS,schneller16alfven,briguglio17mapping}. HAGIS is a perturbative, non-linear wave-particle interaction model which allows the modes to evolve in the presence of EPs, and the EPs to redistribute in phase space due to the interaction with the modes self-consistently. 

The equations of motions in HAGIS are written in Boozer coordinates, thus we assume for the radial coordinates to be related to the normalized poloidal flux $\psi^{1/2} \equiv s \approx r/a$ when transferring the distributions between codes. Calculations are supplied with the equilibrium, fast alpha population and the mode structures introduced in the previous sections. HAGIS uses a $\delta f$ formalism, which allows us to omit the Maxwellian bulk and use the energetic part of the distribution $f_{\rm SD}$ for analysis. The numerically calculated distributions for $t_0 = 0.7\unit{ms}$, $T_f = 3 \unit{eV}$, $t_{\rm N} = 3$ are fitted with the formula shown in eq.~\eref{f_SD.eq} to facilitate implementing the CODION distributions into HAGIS. The alpha particles are represented by $N=10^6$ markers in phase space, which are initialized isotropically in pitch with velocities and positions in the tokamak according to $f_{\rm SD}(r,t_{\rm N} =3)$. Movement along the plasma equilibrium is dictated by their resonant interaction with the calculated modes. The interaction redistributes the particles and transfers energy to the modes, evolving their amplitude. The individual mode growth rates are competing against various damping mechanisms. LIGKA readily provides the damping rates caused by the ion/electron Landau mechanism and radiative damping and are of the order of $\gamma/\omega_{\rm TAE} \approx 0.1\%$. This damping is typically 10 times smaller than reported for TAEs in the pre-disruption phase~\cite{pinches15ions}. The reason is the Landau damping vanishing exponentially as the ion temperature drops to $\approx 1\unit{keV}$~\cite{lauber18IAEA}. 

As the temperature drops further, collisional damping by trapped electrons becomes increasingly important. We use eq.~(2) from Gorelenkov et al. \cite{gorelenkov92collisional} to estimate the collisional damping rate for our disrupting plasma. Damping will be given in $\left[\gamma/\omega\right] = \unit{[s^{-1}/s^{-1}]}$. With the electron-electron collision rate $\nu_e(T)$ and the plasma beta $\beta_e(T)$ its value
\begin{equation*}
\frac{\gamma_e}{\omega_{\rm TAE}} = -2.1 \beta_e \left( \frac{\nu_e}{\omega_{\rm TAE}}\right)^{1/2} \left[ \mathrm{ln} 8 \left( \frac{2 r \omega_{\rm TAE}}{R \nu_e}\right)^{1/2} \right] 
\end{equation*}
reaches $1.2\%$ at $T=60\unit{eV}$ for the $n=8$ mode calculated and only varies slightly for the other modes calculated. Hence we neglect it for $t< 6 t_{\rm N} \simeq 64 \unit{eV}$. Using the resistive MHD code CASTOR (Complex Alfv\'en Spectrum of TORoidal plasmas) \cite{castor} we also calculate fluid damping. Up to a resistivity of $\simeq 0.56 \cdot 10^{-4} \unit{\Omega m}$ which corresponds to $\simeq 6\unit{eV}$ background temperature, the damping is below 1\%, hence will also not be included in our mode evolution simulations.

\begin{figure} [h]
\begin{center}
\includegraphics[width = 0.49\linewidth]{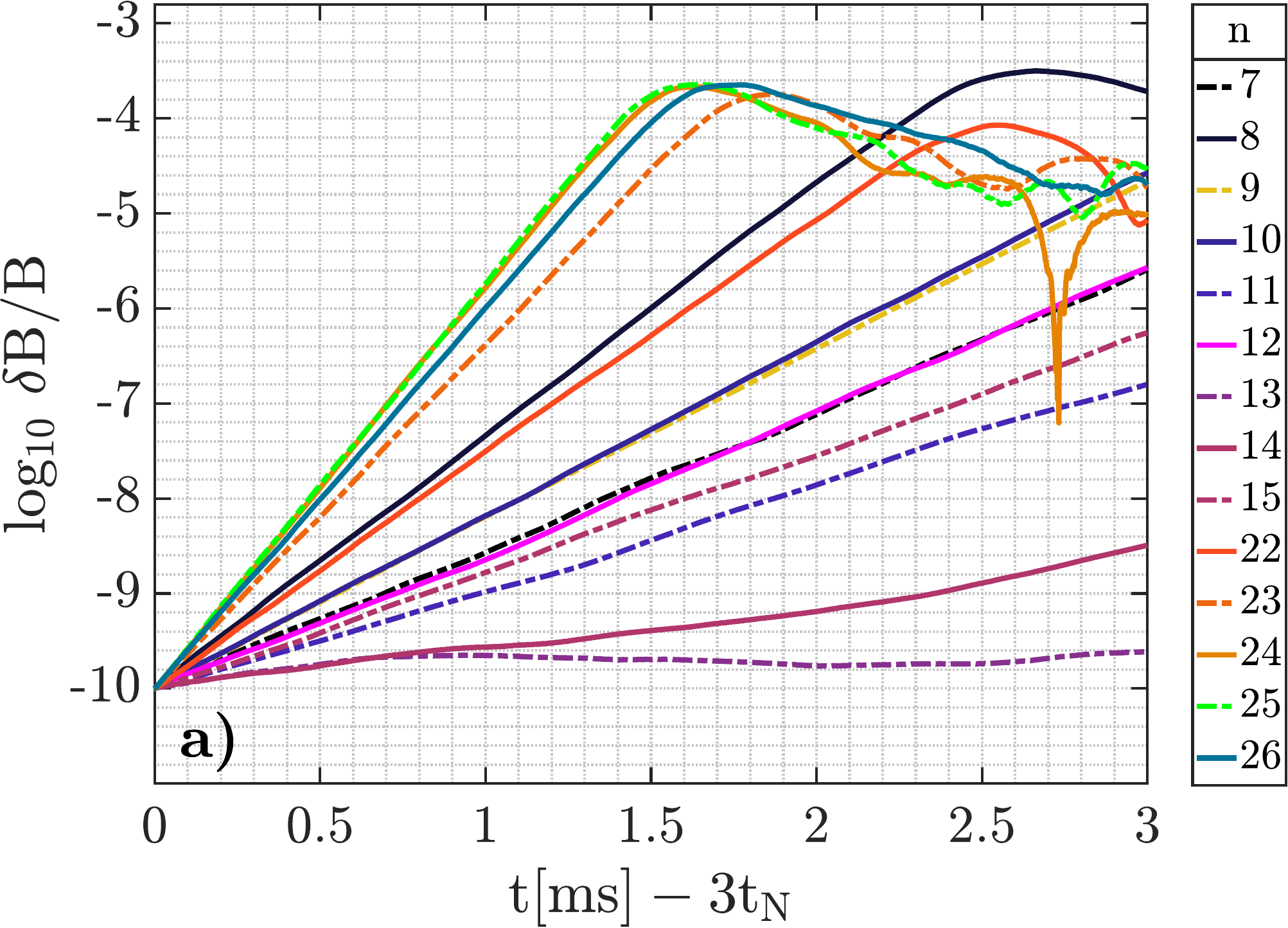} \hfill
\includegraphics[width = 0.49\linewidth]{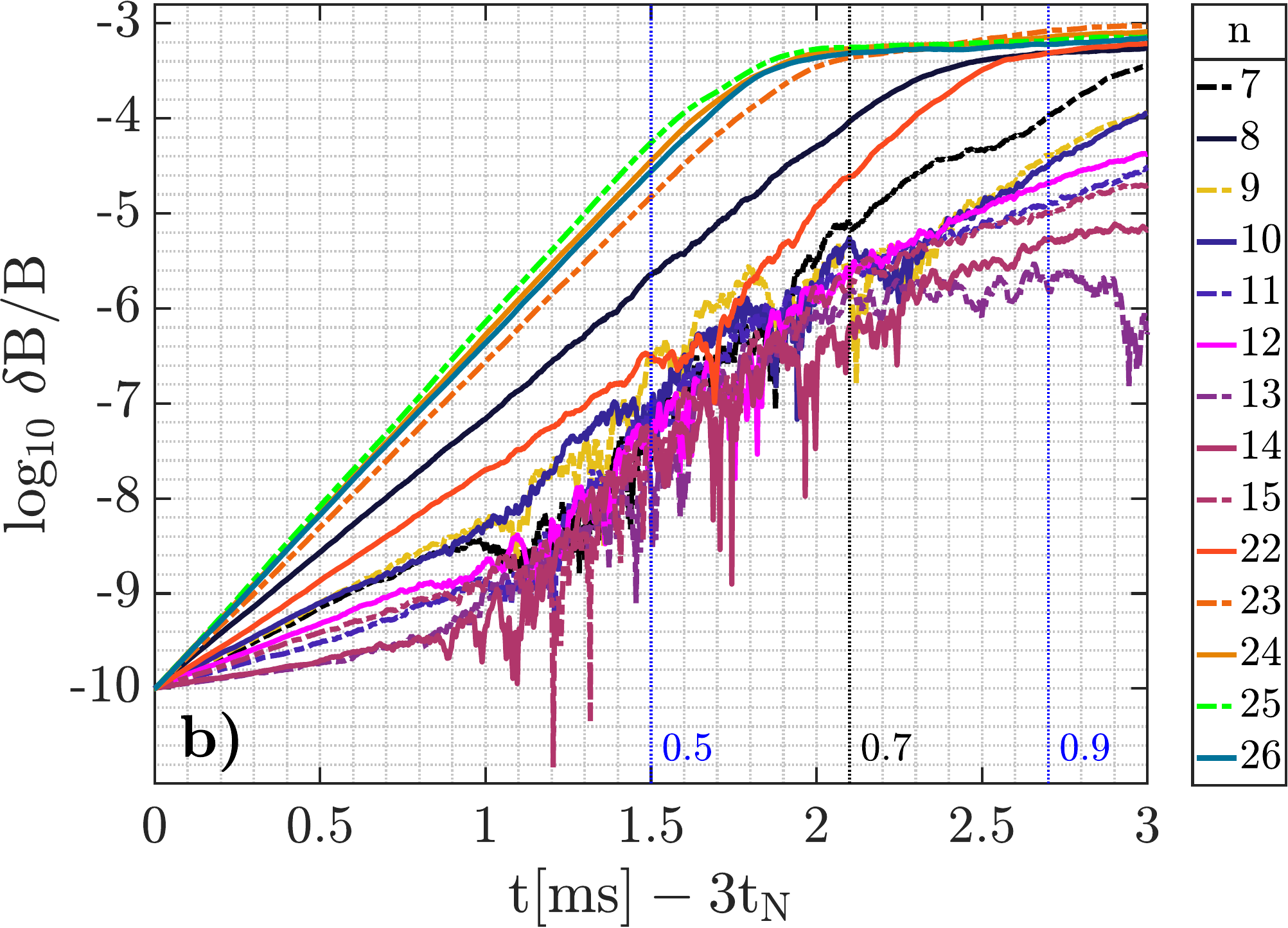}
\caption{Evolution of mode amplitudes $\mathrm{\delta B/B}$ as caused by resonant interaction with the energetic alpha particle population $f_{\rm SD}$, $3t_{\rm N}$ into an ITER disruption in a) single mode and b) multi mode simulation. All available poloidal modes are included. Horizontal lines in b) mark $t_{\rm N} = 6$ for $t_0 = [0.5, 0.7, 0.9]$ from which we extract the individual mode amplitudes for further use.}
\label{fig:mode_evo}
\end{center}
\end{figure}
 
Because the HAGIS model does not include collisional cooling of the driving alpha particles, its driving force only changes due to the redistribution of particles as dictated by their interaction with the modes. The loss of drive due to the thermal quenching however is captured by the CODION simulations. As indicated by \fref{fig:n-T}b) and \fref{fig:Pi_scan}b), the loss of resonance with the Alf\'en velocity (first harmonic) occurs around $t_N = 6$. As this is also the time-point up to which we calculated the collisional damping to be negligible, we restrict the window of mode evolution to $3t_{\rm N}< t < 6t_{\rm N} $. Considering the fact that the dominating (ion Landau) damping mechanism drops exponentially with the temperature, it can be assumed that the mode excitation actually begins earlier than in our computations. Every initial mode amplitude in our simulation is set to $\delta B/B = 10^{-10}$, though previous studies~\cite{pinches15ions,schneller16alfven} have shown TAEs to be only marginally stable in ITER steady state with amplitudes of the order of $\delta B/B = 10^{-5}$ to $10^{-4}$. On the other hand, the HAGIS code is known to overestimate the saturation amplitude due to lack of zonal-flow physics~\cite{schneller13alfv}, these however are typically only relevant at perturbation magnitudes well above the ones discussed in this work.

\Fref{fig:mode_evo}a shows single mode and \fref{fig:mode_evo}b multi mode results. Both are qualitatively different because in multi-mode the energy transfer to waves and the subsequent redistribution of particles may push the modes through multiple resonance regions, as seen by the non-monotonic behavior. In single mode this redistribution may lead to a loss in drive and the damping taking over. The multi-mode simulation is the more realistic one, its linear growth rate of the most pronounced modes $n=23,24,25,26$ (in this time window) saturates at $\delta B/B \approx 0.05\%$ after roughly 2 \unit{ms}, a timescale which is sufficiently short even during a current quench. The effective growth rate in the linear phase for those modes is $\approx 14\%$. A saturation of the $n=8,22$ modes is observed approximately 2.5 \unit{ms} into the simulation. The slowing-down distributions are isotropic in pitch and monotonically decreasing in velocity space, hence the mode drive comes mainly from gradients in real-space and is saturated by its flattening. In \ref{sec:app} we conduct the multi mode evolution with alpha particle densities increased to 130\% and 200\%, resulting in a growth rate in the linear phase of respectively 33\% and 100\%, indicating a high sensitivity. The saturated amplitude however is not strongly affected. 

We want to look at the situation at $t_{\rm N}=6$ into the disruption, since this is right before the avalanching begins (\fref{fig:disr_GO}a) and the energetic particle pressure decays due to collisional cooling (\fref{fig:Pi_scan}b). Depending on the disruption scenario chosen, the amplitudes reached vary, as depicted in \fref{fig:mode_evo}b. The maximum amplitude as well as the root mean square of all mode amplitudes is made note of in table~\ref{table:hagis}. In the next section we simulate the interaction of said modes at their respective amplitudes at $6t_{\rm N}$ for $t_0 = [0.5, 0.7, 0.9]$ with a seed of runaway electrons. 

\begin{table} [htb!]
\centering
\begin{tabular}{c | c | c || c | c}
 $\mathrm{t_0}$ & $\mathrm{max \left(\delta B/B \right)}$ [\%] & $\mathrm{\langle \delta B/B \rangle}$ [\%] & $\mathrm{max \left(\Delta P_\varphi /P_\varphi \right)}$ [\%] & $\mathrm{\langle \Delta P_\varphi /P_\varphi \rangle}$ [\%]\\
 \hline
 0.5 & 0.006 & 0.002 & 3  & 1  \\
 0.7 & 0.056 & 0.027 & 11 & 8  \\
 0.9 & 0.083 & 0.042 & 18 & 10 \\
 \hline 
 0.9 & 0.083 $\cdot$ 1.3 &  0.042 $\cdot$ 1.3  & 39 & 13 \\
 0.9 & 0.083 $\cdot$ 2.0 &  0.042 $\cdot$ 2.0  & 48 & 25 \\ 
\end{tabular}
\caption{HAGIS simulation results: Respectively maximum and average ($\langle \cdot \rangle$) mode amplitudes $\delta B/B$ at $t_{\rm N} = 6$ (\fref{fig:mode_evo}b) and maximum individual (max) and maximum ensemble-averaged ($\langle \cdot \rangle$) RE toroidal momentum changes $\Delta P_\varphi / P_\varphi$ (\fref{fig:pZeta}) caused by said modes after additional $2t_{\rm N}$ to an initial seed of runaway electrons. Additional simulations are run with mode amplitudes of the $t_0=0.9\unit{ms}$ case multiplied by 1.3 and 2 (\fref{fig:pzeta_beta}).}
\label{table:hagis}
\end{table}

%\newpage
\section{Transport of runaway electrons\label{sec:runaway}}
In order to model the interaction of relativistic electrons with Alfv\'en waves, the HAGIS model had to be extended~\cite{papp14interaction}. The derivation of the relativistic equations of motion in Boozer coordinates is provided in \ref{sec:relhagis}. As the runaways are not expected to have a back-reaction to the wave evolution (due to the lack of suitable resonances) the calculations are run in a ``passive'' mode, where only the effect of the presence of the Alfv\'en modes is evaluated on the runaway electron test particles. The mode evolution is disabled and the amplitudes are set to their respective values at ultimately $t_{\rm N} = 6$ into the disruption (\fref{fig:mode_evo}b, table~\ref{table:hagis}).

In order to extract the interaction Green functions, 10000 test particles are launched on a phase space grid in energy, pitch and radial position. The radial region of interest is restricted to $r/a = [0.05 - 0.45]$ and the energies are selected on a logarithmic grid ranging from 10~keV to 30~MeV. For each value of energy, pitch and radial position a total of 25 electrons are distributed evenly on the flux surface with uniform random distribution.  With 5 radial positions, each phase space point is therefore represented by 125 electrons for statistical averaging.

We use the canonical toroidal momentum $P_\varphi$ (equation~\eref{2b}) to quantify the displacement of the test particle orbits. The canonical toroidal momentum is a function of both parallel kinetic momentum and poloidal flux surface, i.e. a change in $P_\varphi$ in a sense represents the contribution to the change in the current profile of the given sub-relativistic test particle. Changes to $P_\varphi$ are dominated by radial transport caused by magnetic perturbation of the modes when no resonant processes are happening. Those are unlikely given that the electron gyro-frequency on axis is $\simeq 150 \unit{GHz}$ and the parallel velocity of the lowest energy electron approximates to $v_{||} \simeq 6 \cdot 10^7 \unit{m/s} \simeq 8 v_A$ (for purely parallel motion).

\begin{figure} [htb!]
\begin{center}
\includegraphics[width = 0.49\linewidth]{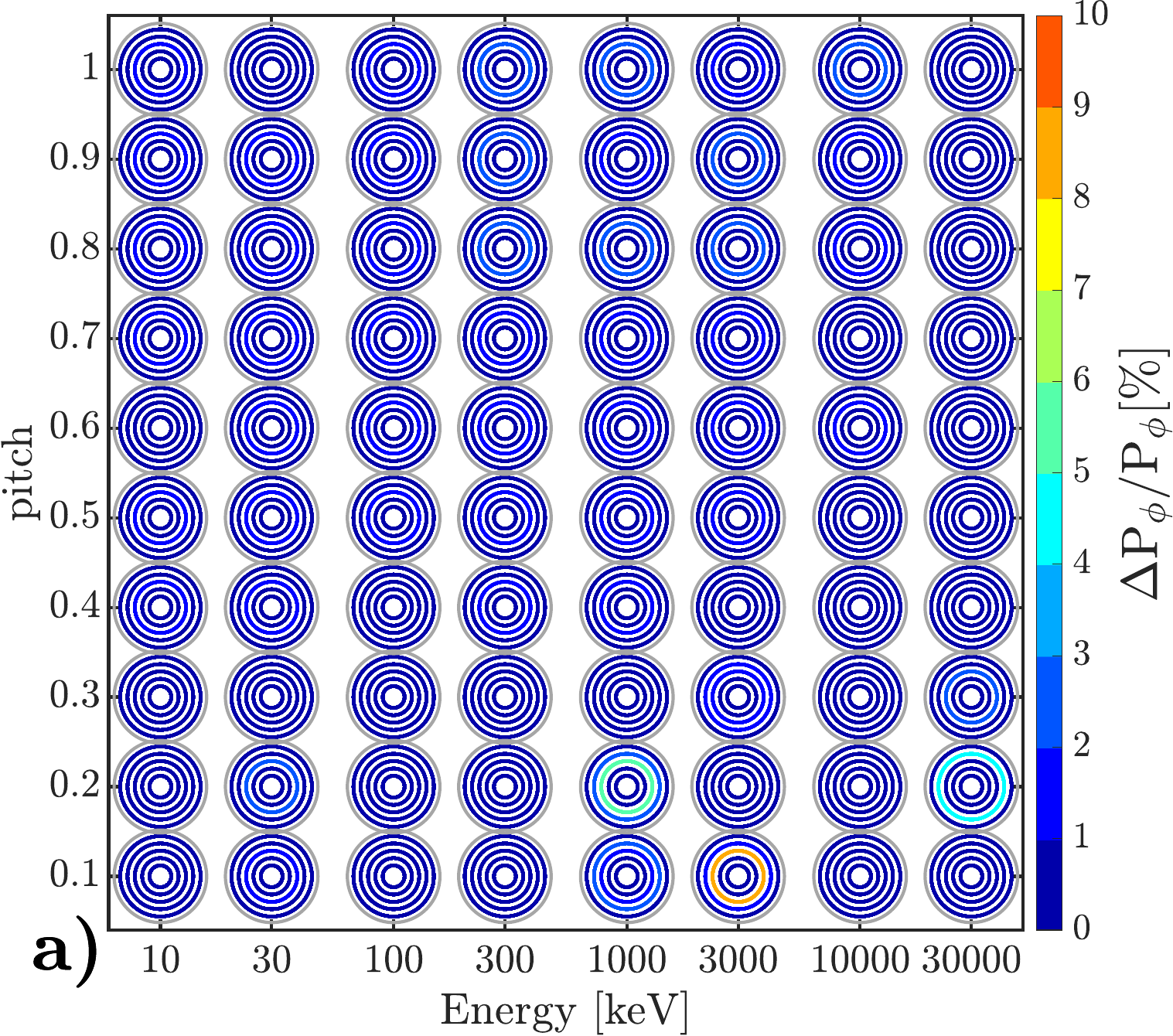} \hfill
\includegraphics[width = 0.49\linewidth]{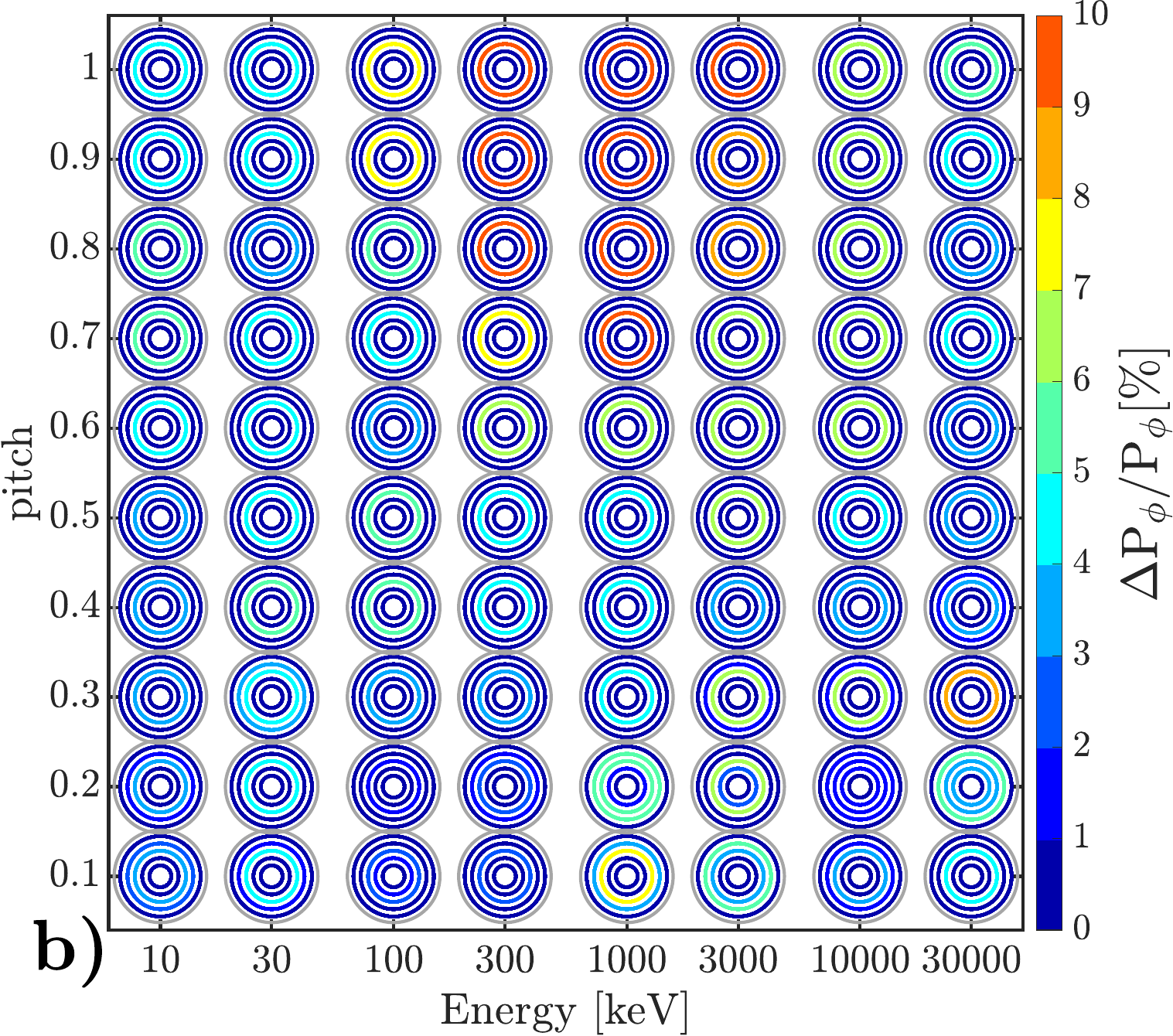}
\caption{Ensemble-averaged relative change (color) in canonical toroidal momentum $P_\varphi$ of the runaway electron test particles as caused by the Alfv\'en modes (a): $t_0 = 0.7\unit{ms}$, b): $t_0 = 0.9\unit{ms}$, table~\ref{table:hagis}). The vertical and horizontal axis represents different initial particle energies and pitches. The radii of the circles represent the radial position of the RE particle in a poloidal cross section of the plasma in the range $r/a=[0.05 - 0.45]$ in steps of 0.1, and bound by a grey circle at $r/a=0.55$.}
\label{fig:pZeta}
\end{center}
\end{figure}

In \fref{fig:pZeta}a we show the results of the passive simulation for the amplitudes extracted for the chosen $t_0=0.7\unit{ms}$ case after $2t_{\rm N}$ integration time corresponding to 125 poloidal turns of each runaway particle. Each circle represents a radial starting position and its color shows the overall change in $P_\varphi$, ensemble-averaged over the 25 REs that started on this flux surface. We observe minor changes of few \% to $P_\varphi$ in few selected phase-space positions. 
However, when we take the mode amplitudes reached further into the thermal quench (\fref{fig:mode_evo}b, $2.7\unit{ms}$, corresponding to ultimately $6 t_{\rm N}$ for $t_0 = 0.9\unit{ms}$), this picture changes significantly (\fref{fig:pZeta}b). As expected from the mode structures, the inner- and outermost particles initiated are not affected by the TAEs, however, the change is up to 10\% for REs localized in between. The effect is stronger for higher particle pitches but applies for a wide range in the phase space. 

Average/maximum values for the change in $P_\varphi$ are collected in table~\ref{table:hagis} for those simulations, including a reference run for the $t_0 = 0.5\unit{ms}$ case. In addition, as a sensitivity scan we carried out simulations with 130\% and 200\% individual mode amplitude of the $t_0 = 0.9 \unit{ms}$ case (\fref{fig:pzeta_beta}, table~\ref{table:hagis}). This fulfils the purpose of both a numerical sensitivity scan as well as it considers the experimental possibility of external amplitude amplification. At twice the amplitudes ($\max(\delta B/B) \approx 0.17\%$) the average displacement on a flux surface is found to be as high as 25\% for various points in initial phase space, including the most dangerous MeV REs. 

These simulations do not evaluate the runaway electron dynamics, but serve as an indication to the possibility to transport runaways via alpha particle driven modes. Recent studies~\cite{svensson20magper} suggest that the perturbation amplitudes and particle displacement caused by the modes discussed in this paper can lead to runaway avalanche mitigation (or even suppression). If the core transport is enhanced by for example alpha-driven modes, runaways are more easily transported to the edge, where other methods, such as Resonant Magnetic Perturbations~\cite{papp12rmp,papp15magper} could further aid the removal of runaways.

\section{Summary and outlook}
In this paper we simulate unmitigated disruptions in burning ITER plasmas and we show that alpha particles remain sufficiently energetic during the current quench to drive Alfv\'en modes sustained by the current quench equilibrium. In turn these modes can enhance the transport of the runaway electron seed prior to the onset of significant induced electric field, preceding the runaway avalanche.

The disruptions considered are characterized by an exponential temperature decay time $t_0=[0-1]\unit{ms}$, and we have conducted a parameter scan for the final temperature and speed of the thermal quench using the GO code. The output of these simulations was used in the Fokker-Planck solver CODION to track the collisional cooling of the alpha particle distribution. We find that the energetic tail is sustained into the current quench. Throughout the parameter range, the evolution of the simulated core fast particle pressure shows a general similarity up to $t_{\rm N} \equiv t/t_0 = 3$. The simulated current and pressure profiles are used as input to construct equilibria using the VMEC code. The Alfv\'en continuum of these equilibria is analysed using LIGKA, showing the possible existence of core-localized TAEs.

Using HAIGS we have calculated the growth and saturation amplitudes of these TAEs, driven by the energetic alpha particles. The modes reach a saturation level of up to $\delta B / B \approx 0.1\%$ for cases of $t_0 > 0.7 \unit{ms}$. Runaway electron test particles were used in the HAGIS simulations to analyse the impact of these TAEs on runaway transport. The onset and saturation of the modes occurs before the rise of the electric field induced in the current quench, before the start of significant runaway avalanche. We show that the runaway electrons can be subjected to significant displacement due to the presence of the TAEs, which may contribute to a reduction of the runaway electron avalanche. The parameter and sensitivity scans indicate that slower thermal quenches are beneficial from the perspective of the studied phenomenon, and that there is a high sensitivity of runaway transport to mode amplitude.

All of our simulations have considered good confinement of alpha particles after the thermal quench. We argue that this is a conservative estimate, i.e. when the breakup of magnetic surfaces is bad enough to lead to alpha particle losses, the runaway electrons - which have higher thermal speeds - are likely to get lost as well. A rising runaway current in the current quench may in fact contribute to alpha particle confinement and in turn the drive of TAEs. This paper analyses the ``worst case scenario'' of unmitigated disruptions. Future studies will have to be conducted to evaluate the possibility of alpha particle drive of Alfv\'enic instabilities in ITER disruptions mitigated by e.g. shattered pellet injection. or other means. 

In conclusion, natural disruptions in D-T ITER plasmas may be able to provide a natural mechanism that contributes to runaway avalanche suppression. Following this proof-of-principle paper, further studies are necessary to identify the optimal disruption scenario which maximizes runaway transport, and the self-consistent evaluation of runaway dynamics will also be necessary.

\section*{Acknowledgments}
The authors are grateful to E.~Strumberger, Th.~W.~Hayward-Schneider, M.~Schneller and S.~Newton for fruitful discussions.
Ph.~W. Lauber acknowledges support from the Enabling
research projects NAT (CfP-AWP17-ENR-MPG-01) and MET (CfP-AWP19-ENR-01-ENEA-05).
This work has been carried out within the framework of the EUROfusion Consortium and has received funding from the Euratom research and training programme 2014-2018 and 2019-2020 under grant agreement No 633053. The views and opinions expressed herein do not necessarily reflect those of the European Commission. 

\begin{appendix}
\section{Mode amplitude sensitivity scan\label{sec:app}}

\begin{figure} [htb!]
\begin{center}
\includegraphics[width=0.49\linewidth]{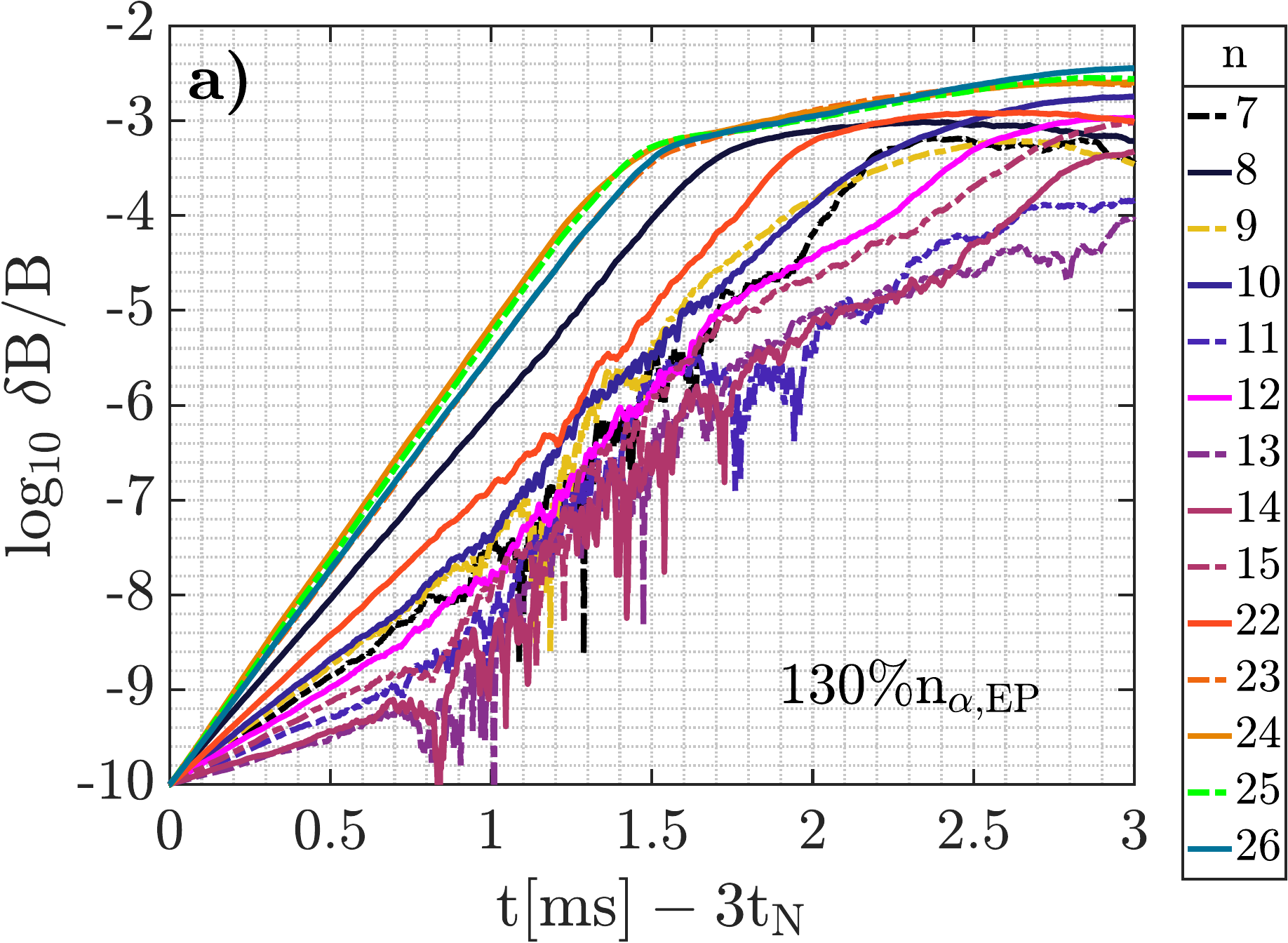} \hfill
\includegraphics[width=0.49\linewidth]{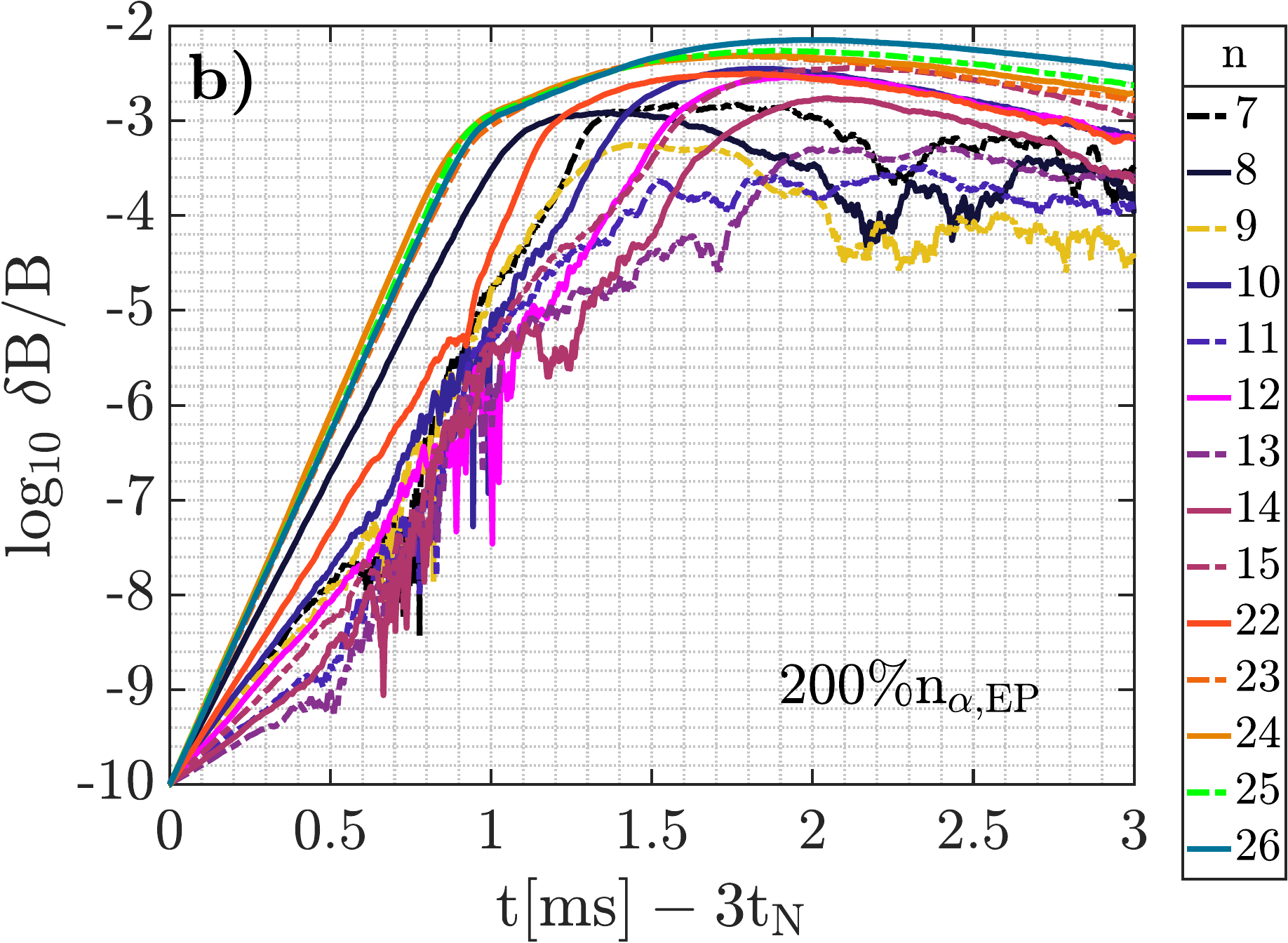}
\caption{Multi mode amplitude evolution of available TAEs (including poloidal harmonics) during an ITER disruption as caused by an energetic alpha particle population. Its originally calculated density $3t_{\rm N}$ into the disruption is increased to a) 130\% and b) 200\% to establish a sensitivity estimation of \fref{fig:mode_evo}b.}
\label{fig:sensitivity_mode}
\end{center}
\end{figure}

%\subsection{\label{sec:sensitivity_mode} Sensitivity scan: Multi mode evolution} 
\Fref{fig:sensitivity_mode} presents a sensitivity scan of the multi mode amplitude growth caused by the energetic alpha particle population as computed by HAGIS. While the original alpha particle density causes a growth rate of (at max) 14\% in the linear phase, this growth rate increases to 33\% (a) and 100\% (b) for respectively 130\% and 200\% original density values. Though saturated earlier, the amplitudes reached after the linear phase are approximately the same. HAGIS is valid up to $\delta B/B \approx 1\%$.

\begin{figure}
\begin{center}
\includegraphics[width = 0.49\linewidth]{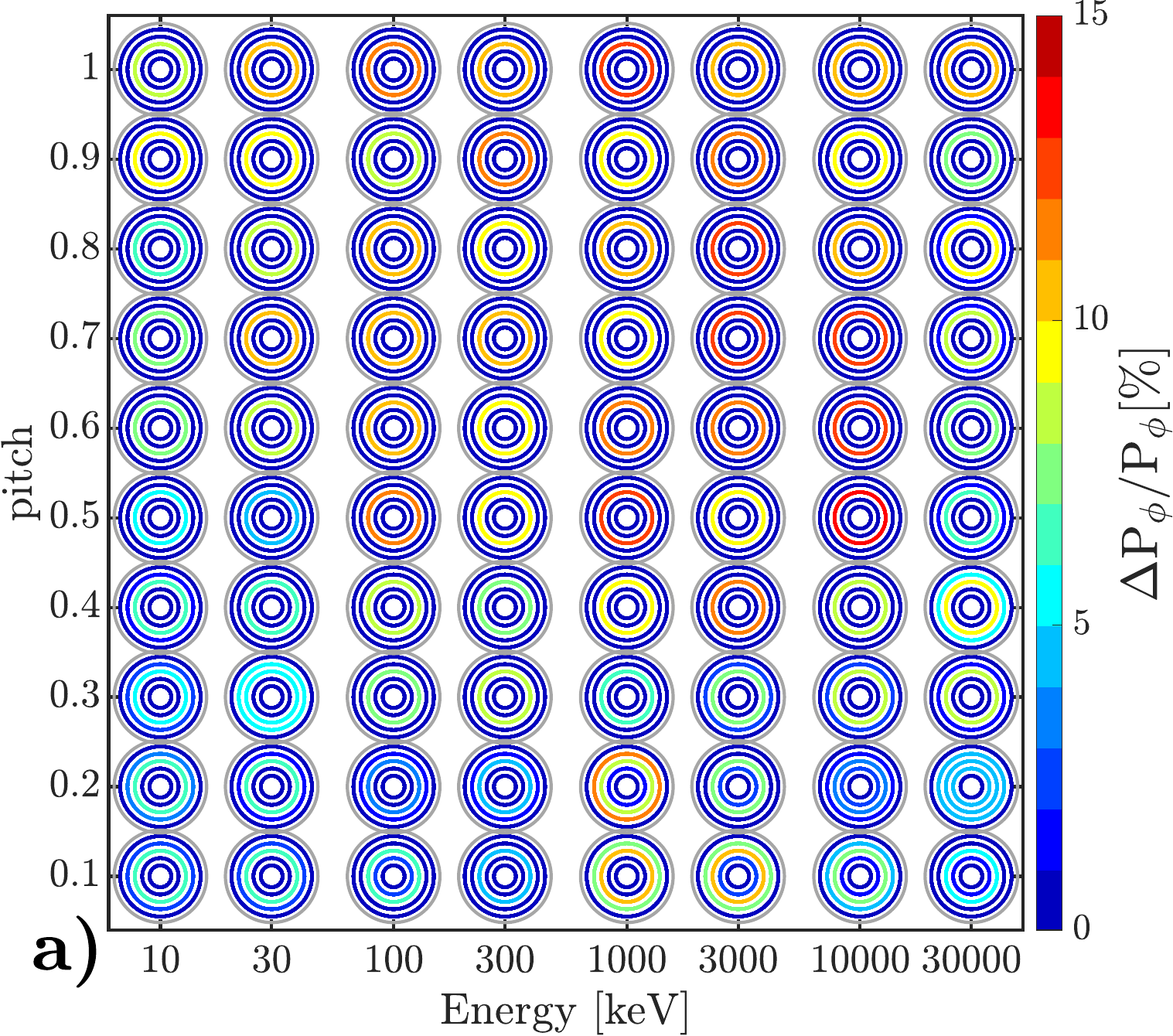} \hfill
\includegraphics[width = 0.49\linewidth]{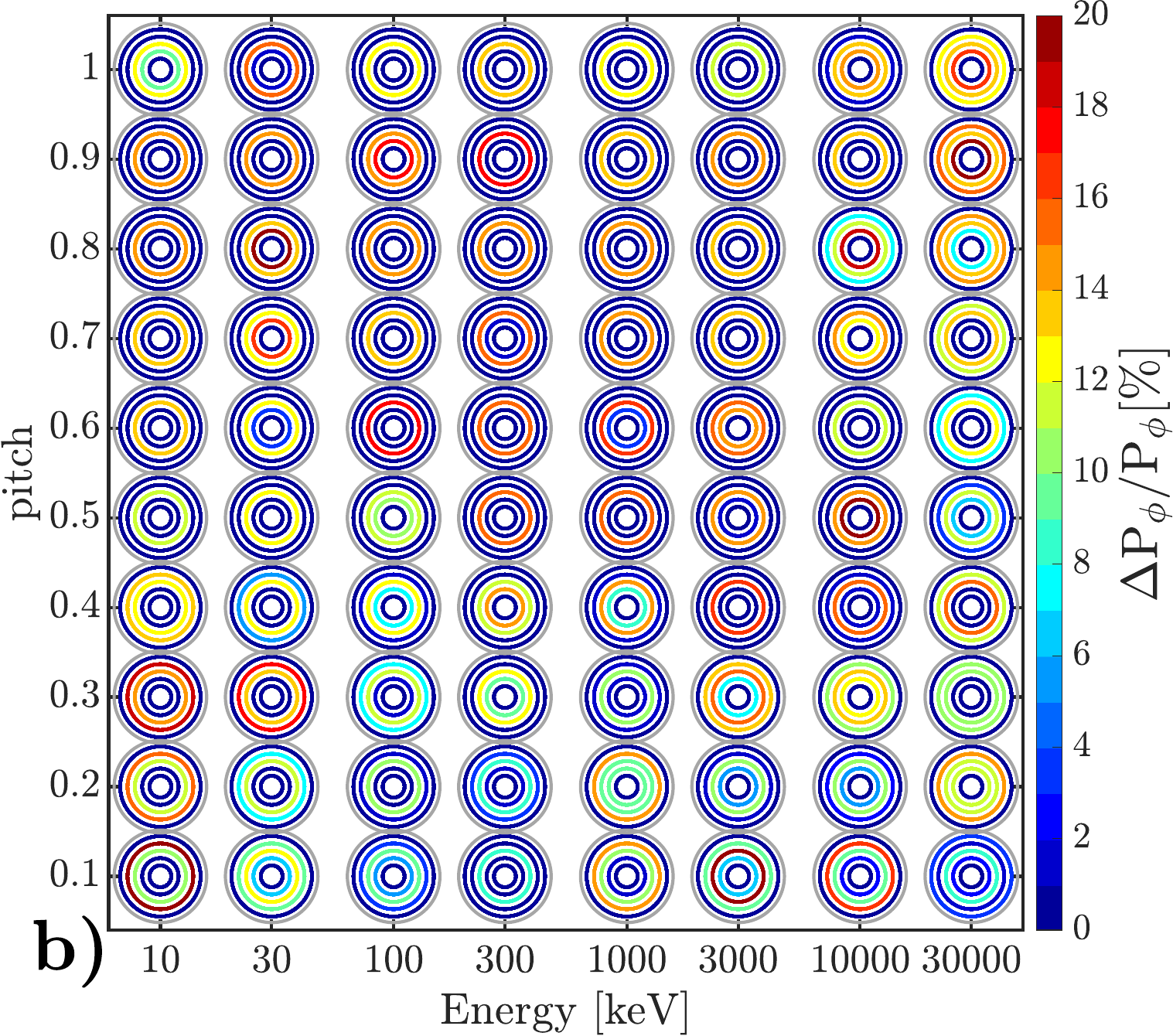}
\caption{Ensemble-averaged relative change (color) in canonical toroidal momentum $P_\varphi$ of the runaway electron test particles as caused by amplified (a): $max(\delta B/B) = 0.108\%$ and $\langle \delta B/B \rangle = 0.54 \%$ , b): $max(\delta B/B) = 0.166\%$ and $\langle \delta B/B \rangle = 0.84 \%$) Alfv\'en modes (table~\ref{table:hagis}, case $t_0 = 0.9\unit{ms}$, amplified). The vertical and horizontal axis represents different particle energies and pitches. The radii of the circles represent the radial position of the RE particle in a poloidal cross section of the plasma in the range $r/a=[0.05 - 0.45]$ in steps of 0.1, and bound by a grey circle at $r/a=0.55$. Note the difference in the maximum for the color scale.}
\label{fig:pzeta_beta}
\end{center}
\end{figure}

%\subsection{\label{sec:sensitivity_transport} Sensitivity scan: Transport of runaway electrons}
In \fref{fig:pzeta_beta} a scan is conducted where we take the respective $t_0 = 0.9\unit{ms}$ mode amplitudes at a) 130\% and b) 200\% (see table~\ref{table:hagis}). This serves both as a sensitivity scan and a means to explore the influence of additional external amplitude amplification. We observe a significant increase of runaway displacement as the TAE amplitude is raised, which suggests that a disruption scenario can be optimized to maximize runaway seed transport.

%\newpage
\section{\label{sec:relhagis} Relativistic equations of motion for runaway electrons in HAGIS}

In this appendix, we derive the relativistic equations of motion for runaway electrons in HAGIS. The numerical implementation has been verified~\cite{papp14interaction} through extensive comparisons with the ANTS code~\cite{papp11runaway,papp11iter,papp12iter,papp15energetic} by tracking the same test particle ensembles in the same background equilibria. The following is a relativistic extension to the derivations in the PhD thesis of S.~Pinches~\cite{pinches96phd}. The constants used are $e$ for the electric charge, $c$ for the speed of light, $m$ for the particle mass.
The relativistic guiding-centre Lagrangian reads (in CGS units)\cite{cary09hamiltonian}
\begin{eqnarray*}
L(\vect{x},\vect{\dot{x}},t) &=& \left[ \frac{e}{c} \vect{A}(\vect{x},t) + \frac{p_{\para}}{B} \vect{B}(\vect{x},t)\right] \cdot \vect{\dot{x}} + \frac{mc}{e} \mu \dot{\vartheta} - H_{gc} 
\end{eqnarray*}
with the relativistic guiding-centre Hamiltonian
\begin{eqnarray*}
H_{gc} &=& \gamma mc^2 +e\phi(\vect{x},t) 
\end{eqnarray*}
and the relativistic factor
\begin{eqnarray*}
\gamma &=& \sqrt{1+\frac{2}{mc^2} \mu B + \frac{p_{\para}^2}{m^2c^2}}.
\end{eqnarray*}
$\vect{A}$ and $\phi$ are the vector and electric potential, respectively, $\vect{x}$ is the guiding-centre position, $\vartheta$ is the gyroangle, $B$ the magnetic field strength, and the parallel momentum $p_{\para}$ and the magnetic moment $\mu$ are defined as follows:
\begin{eqnarray*}
p_{\para}&=& m \gamma \vect{v} \cdot \vect{b}, \\
\mu &=& m \gamma^2 \frac{\vect{v}_{\perp}^2}{2B}
\end{eqnarray*} 
where $\vect{b}$ is the unit vector along the magnetic field and $\vect{v}$ is the particle velocity vector. 
In Boozer coordinates, the magnetic field can be represented as
\begin{eqnarray*}
\vect{B} &=& I(\psi_p) \nabla \theta + g(\psi_p) \nabla \varphi + \delta(\psi_p,\theta) \nabla \psi_p,
\end{eqnarray*}
and the vector potential as
\begin{eqnarray*}
\vect{A} &=& \psi_t \nabla \theta -\psi_p \nabla \varphi.
\end{eqnarray*}
$\psi_t$ and $\psi_p$ are the toroidal and poloidal fluxes, $I$ and $g$ are the toroidal and poloidal currents, $\delta$ the radial covariant component of $\vect{B}$, and $\theta$ and $\varphi$ the poloidal and toroidal angles, respectively. 
Using Boozer coordinates for the guiding-centre Lagrangian, we find
\begin{eqnarray*}
L&=& \left(\frac{e}{c} \vect{A} + \frac{p_{\para}}{B} \vect{B} \right) \cdot \dot{\vect{x}} +\frac{mc}{e} \mu \dot{\vartheta} -H_{gc} \\
&=& \left(\frac{e}{c} \left(\psi_t \nabla \theta -\psi_p \nabla \varphi \right)+\frac{p_{\para}}{B} \left(\delta \nabla \psi_p + I \nabla \theta +g \nabla \varphi \right) \right) \cdot \dot{\vect{x}} + \frac{mc}{e} \mu \dot{\vartheta} -H_{gc} \\
&=& \left(\frac{e}{c} \psi_t + \frac{p_{\para}I}{B} \right) \dot{\theta} + \left(\frac{p_{\para}}{B} g - \frac{e}{c} \psi_p \right) \dot{\varphi} +\frac{p_{\para}}{B} \delta \dot{\psi}_p + \frac{mc}{e} \mu \dot{\vartheta} -H_{gc}.
\end{eqnarray*}
In order to achieve the natural Lagrarian form and read the canonical momenta straight away, we have to reabsorb $\dot{\psi}_p$. This is done using the same argumentation as on p.47/48 of S.~Pinches PhD thesis~\cite{pinches96phd} for the derivation for the non-relativistic equations: we transform the guiding center velocity $\dot{x} \rightarrow \dot{x} + w$ with
\begin{eqnarray*}
\vect{w} &=& - \frac{\left(\frac{p_{\para}}{B} \delta + \frac{e}{c} \tilde{A}_{\psi_p}\right) \dot{\psi}_p}{\frac{e}{c} \vect{A}\cdot\vect{B} +\frac{p_{\para}}{B} B^2} \vect{B}.
\end{eqnarray*}
We now add perturbations to both potentials of the form
\begin{eqnarray*}
\vect{A}(\vect{x},t) &=& \tilde{A}_{\psi_p} \nabla \psi_p +\tilde{A}_{\theta} \nabla \theta + \tilde{A}_{\varphi} \nabla \varphi \qquad \qquad \text{and} \qquad \qquad \tilde{\phi}=\tilde{\phi}(\vect{x},t),
\end{eqnarray*}
and are now able to express the canonical momenta for the set of canonical variables $(\theta,\varphi,\vartheta)$ as
\begin{eqnarray}
P_{\theta} &=& \frac{p_{\para}}{B} I + \frac{e}{c} \psi_t + \frac{e}{c} \tilde{A}_{\theta} \label{2a}\\
P_{\varphi} &=& \frac{p_{\para}}{B} g - \frac{e}{c} \psi_p + \frac{e}{c} \tilde{A}_{\varphi} \label{2b}\\
\nonumber P_{\vartheta} &=& \frac{mc}{e} \mu.
\end{eqnarray}

\subsection{Equations of motion}
Combining canonical momenta \eqref{2a} and \eqref{2b} yields
\begin{eqnarray*}
g \left(P_{\theta} -\frac{e}{c} \psi_t -\frac{e}{c} \tilde{A}_{\theta} \right)&=&I \left(P_{\varphi} + \frac{e}{c} \psi_p - \frac{e}{c} \tilde{A}_{\varphi}\right).
\end{eqnarray*}
We differentiate with respect to $\theta,\varphi, P_{\theta}$ and $P_{\varphi}$ and using $q = \partial \psi_t / \partial \psi_p$ to obtain
\begin{eqnarray*}
\frac{d}{d \theta}: && g^{\prime} \frac{\partial \psi_p}{\partial \theta} \frac{p_{\para} I}{B} + g \left( -q \frac{e}{c} - \frac{e}{c} \left(\frac{\partial \tilde{A}_{\theta}}{\partial \theta}+ \tilde{A}_{\theta}^{\prime} \frac{\partial \psi_p}{\partial \theta} \right) \right) \\
&& -I^{\prime} \frac{\partial \psi_p}{\partial \theta} \frac{p_{\para} g}{B} -I \left( \frac{e}{c} -\frac{e}{c} \left(\frac{\partial\tilde{A}_{\varphi}}{\partial \theta}+\tilde{A}_{\varphi}^{\prime} \frac{\partial \psi_p}{\partial \theta} \right) \right) =0 \\
\Leftrightarrow && \frac{\partial \psi_p}{\partial \theta} = \frac{1}{D_r} \left(I \frac{\partial \tilde{A}_{\varphi}}{\partial \theta} - g \frac{\partial \tilde{A}_{\theta}}{\partial \theta} \right)
\end{eqnarray*}
where
\begin{eqnarray*}
D_r &=& \frac{c}{e} \frac{p_{\para}}{B} (I^{\prime} g -I g^{\prime}) +I +qg +g \tilde{A}_{\theta}^{\prime} -I \tilde{A}_{\varphi}^{\prime}.
\end{eqnarray*}
Analogously,
\begin{eqnarray*}
\frac{\partial \psi_p}{\partial \varphi} &=& \frac{1}{D_r} \left(I \frac{\partial \tilde{A}_{\varphi}}{\partial \varphi} - g \frac{\partial \tilde{A}_{\theta}}{\partial \varphi} \right).
\end{eqnarray*}
Thus the next differentiation becomes
\begin{eqnarray*}
\frac{d}{d P_{\theta}}: && g^{\prime} \frac{\partial \psi_p}{\partial P_{\theta}}\frac{p_{\para} I}{B} + g \left(1-\frac{e}{c} q \frac{\partial \psi_p}{\partial P_{\theta}} -\frac{e}{c} \tilde{A}_{\theta}^{\prime} \frac{\partial \psi_p}{\partial P_{\theta}} \right. \\
&&\left. - I^{\prime} \frac{\partial \psi_p}{\partial P_{\theta}}  \frac{p_{\para} g}{B} -I \left( \frac{e}{c} - \frac{e}{c} \tilde{A}_{\varphi}^{\prime} \frac{\partial \psi_p}{\partial P_{\theta}} \right)\right)=0 \\
\Leftrightarrow && \frac{\partial \psi_p}{\partial P_{\theta}} = \frac{c}{e} \frac{g}{D_r}
\end{eqnarray*}
and
\begin{eqnarray*}
\frac{\partial \psi_p}{\partial P_{\varphi}} &=& - \frac{c}{e}\frac{I}{D_r}
\end{eqnarray*}
when performing the same operation with $d/dP_\varphi$.
Differentiation of $p_{\para}/B$ with respect to $\theta, \varphi, P_{\theta}$ and $P_{\varphi}$ yields
\begin{eqnarray*}
\frac{d}{d \theta}: & 0 = \frac{\partial}{\partial \theta}\left(\frac{p_{\para}}{B}\right) I + \frac{p_{\para}}{B} I^{\prime} \frac{\partial \psi_p}{\partial \theta} + \frac{e}{c}q \frac{\partial \psi_p}{\partial \theta} + \frac{e}{c} \left(\frac{\partial \tilde{A}_{\theta}}{\partial \theta} + \tilde{A}_{\theta}^{\prime} \frac{\partial \psi_p}{\partial \theta} \right) \\
\Leftrightarrow & \frac{\partial}{\partial \theta}\left(\frac{p_{\para}}{B}\right) = -\frac{1}{ID_r} \left( \left(\frac{p_{\para}}{B} I^{\prime} + \frac{e}{c} q + \frac{e}{c} \tilde{A}_{\theta}^{\prime} \right) \left(I \frac{\partial \tilde{A}_{\varphi}}{\partial \theta} -g \frac{\partial \tilde{A}_{\theta}}{\partial \theta} \right) +\frac{e}{c} \frac{\partial \tilde{A}_{\theta}}{\partial \theta} D_r \right) \\
&=-\frac{1}{D_r} \left(\frac{\partial \tilde{A}_{\varphi}}{\partial \theta} \left( \frac{p_{\para}}{B} I^{\prime} + \frac{e}{c} q + \frac{e}{c} \tilde{A}_{\theta}^{\prime} \right) + \frac{\partial \tilde{A}_{\theta}}{\partial \theta} \left(-\frac{p_{\para}}{B} g^{\prime} + \frac{e}{c} - \frac{e}{c} \tilde{A}_{\varphi}^{\prime} \right) \right) \\
\Leftrightarrow & \frac{\partial}{\partial \theta} \left(\frac{p_{\para}}{B} \right) = \frac{1}{D_r} \left( \frac{\partial \tilde{A}_{\theta}}{\partial \theta} \left(\frac{p_{\para}}{B} g^{\prime} -\frac{e}{c} + \frac{e}{c} \tilde{A}_{\varphi}^{\prime} \right) - \frac{\partial \tilde{A}_{\varphi}}{\partial \theta} \left(\frac{p_{\para}}{B} I^{\prime} + \frac{e}{c} q + \frac{e}{c} \tilde{A}_{\theta}^{\prime} \right) \right). 
\end{eqnarray*}
Analogously
\begin{eqnarray*}
\frac{\partial}{\partial \varphi} \left(\frac{p_{\para}}{B} \right) &=& \frac{1}{D_r} \left(\frac{\partial \tilde{A}_{\theta}}{\partial \varphi} \left(\frac{p_{\para}}{B} g^{\prime} -\frac{e}{c} +\frac{e}{c} \tilde{A}_{\varphi}^{\prime} \right) - \frac{\partial \tilde{A}_{\varphi}}{\partial \varphi} \left( \frac{p_{\para}}{B} I^{\prime} + \frac{e}{c} q + \frac{e}{c} \tilde{A}_{\theta}^{\prime} \right) \right).
\end{eqnarray*}

\begin{eqnarray*}
\frac{d}{d P_{\theta}}: 1 &=& \frac{\partial}{\partial P_{\theta}} \left(\frac{p_{\para}}{B} \right) I + \frac{p_{\para}}{B} I^{\prime} \frac{\partial \psi_p}{\partial P_{\theta}} + \frac{e}{c}q\frac{\partial \psi_p}{\partial P_{\theta}}+\frac{e}{c} \tilde{A}_{\theta}^{\prime} \frac{\partial \psi_p}{\partial P_{\theta}} \\
\Leftrightarrow && \frac{\partial}{\partial P_{\theta}} \left(\frac{p_{\para}}{B}\right) = \frac{1}{ID_r} \left(D_r-\left(\frac{p_{\para}}{B} I^{\prime} +\frac{e}{c} q +\frac{e}{c} \tilde{A}_{\theta}^{\prime} \right) \frac{c}{e} g \right) \\
&&=\frac{1}{D_r} \left(1-\tilde{A}_{\varphi}^{\prime} -\frac{c}{e} \frac{p_{\para}}{B} g^{\prime} \right)
\end{eqnarray*}
and
\begin{eqnarray*}
\frac{\partial}{\partial P_{\varphi}} \left(\frac{p_{\para}}{B} \right) &=&\frac{1}{D_r} \left(q+\tilde{A}_{\theta}^{\prime} +\frac{c}{e} \frac{p_{\para}}{B} I^{\prime} \right)
\end{eqnarray*}
Combining above expressions for the equations of motion
\begin{eqnarray*}
\frac{\partial H}{\partial P_{\theta}} &=& \dot{\theta} = \frac{\partial H}{\partial \psi_p} \frac{\partial \psi_p}{\partial P_{\theta}} + \frac{\partial H}{\partial \left(\frac{p_{\para}}{B}\right)} \frac{\partial \left(\frac{p_{\para}}{B}\right)}{\partial P_{\theta}} \\
\frac{\partial H}{\partial \psi_p} &=& \frac{B^{\prime}}{\gamma} \left(\mu+\frac{B}{m} \left(\frac{p_{\para}}{B}\right)^2\right)+e\phi^{\prime} \\
\frac{\partial H}{\partial \left(\frac{p_{\para}}{B} \right)} &=& \frac{B^2}{\gamma m} \frac{p_{\para}}{B}
\end{eqnarray*}
yields 
\begin{eqnarray*}
\Leftrightarrow \dot{\theta} &=& \frac{1}{D_r}\left[ \left(\frac{B^{\prime}}{\gamma} \left(\mu +\frac{B}{m} \left(\frac{p_{\para}}{B}\right)^2\right) +e\phi^{\prime} \right) \frac{c}{e} g + \frac{B^2}{\gamma m} \frac{p_{\para}}{B} \left(1-\tilde{A}_{\varphi}^{\prime} -\frac{c}{e} \frac{p_{\para}}{B} g^{\prime} \right) \right] \\
\dot{\varphi} &=& \frac{1}{D_r} \left[-\left(\frac{B^{\prime}}{\gamma} \left(\mu +\frac{B}{m} \left(\frac{p_{\para}}{B}\right)^2\right) +e\phi^{\prime} \right) \frac{c}{e} I + \frac{B^2}{\gamma m} \frac{p_{\para}}{B} \left(q+\tilde{A}_{\theta}^{\prime} + \frac{c}{e} \frac{p_{\para}}{B} I^{\prime} \right) \right] \\
\dot{P}_{\theta} &=& - \frac{1}{\gamma} \left(\mu +\frac{B}{m} \left(\frac{p_{\para}}{B}\right)^2\right) \frac{\partial B}{\partial \theta} - e \frac{\partial \phi}{\partial \theta}  - \frac{1}{D_r} \left(\frac{B^{\prime}}{\gamma}\left(\mu+\frac{B}{m} \left(\frac{p_{\para}}{B}\right)^2 +e\phi^{\prime} \right)\left(I \frac{\partial \tilde{A}_{\varphi}^{\prime}}{\partial \theta} - g \frac{\partial \tilde{A}_{\theta}}{\partial \theta} \right) \right.\\
&&\left.-\frac{1}{D_r} \frac{B^2}{\gamma m} \frac{p_{\para}}{B} \left(\left(\frac{p_{\para}}{B} g^{\prime} - \frac{e}{c} +\frac{e}{c} \tilde{A}_{\varphi}^{\prime} \right) \frac{\partial \tilde{A}_{\theta}}{\partial \theta} - \left(\frac{p_{\para}}{B} I^{\prime} + \frac{e}{c} q + \frac{e}{c} \tilde{A}_{\theta}^{\prime} \right)\frac{\partial \tilde{A}_{\varphi}^{\prime}}{\partial \theta} \right)\right)\\
\dot{P}_{\varphi} &=& -\frac{1}{\gamma}\left(\mu + \frac{B}{m} \left(\frac{p_{\para}}{B}\right)^2\right) \frac{\partial B}{\partial \varphi} - e \frac{\partial \phi}{\partial \varphi} -\frac{1}{D_r}\left(\frac{B^{\prime}}{\gamma} \left(\mu + \frac{B}{m} \left(\frac{p_{\para}}{B}\right)^2 +e\phi^{\prime} \right)\left(I \frac{\partial \tilde{A}_{\varphi}}{\partial \varphi} -g \frac{\tilde{A}_{\theta}}{\partial \varphi} \right)\right. \\
&&\left. -\frac{1}{D_r}\frac{B^2}{\gamma m} \frac{p_{\para}}{B} \left( \left(\frac{p_{\para}}{B} g^{\prime} -\frac{e}{c} +\frac{e}{c} \tilde{A}_{\varphi}^{\prime} \right) \frac{\partial \tilde{A}_{\theta}}{\partial \varphi} - \left(\frac{p_{\para}}{B} I^{\prime} +\frac{e}{c} q +\frac{e}{c} \tilde{A}_{\theta}^{\prime} \right) \frac{\partial \tilde{A}_{\varphi}}{\partial \varphi} \right) \right).
\end{eqnarray*}
Furthermore the change in poloidal flux can now be written as 
\begin{eqnarray*}
\dot{\psi}_p &=& \frac{\partial \psi_p}{\partial \theta} \dot{\theta} +\frac{\partial \psi_p}{\partial \varphi} \dot{\varphi} +\frac{\partial \psi_p}{\partial P_{\theta}} \dot{P}_{\theta} + \frac{\partial \psi_p}{\partial P_{\varphi}} \dot{P}_{\varphi} \\
&=& \frac{1}{D_r} \left[ \left(I \frac{\partial \tilde{A}_{\varphi}}{\partial \theta} - g \frac{\partial \tilde{A}_{\theta}}{\partial \theta} \right) \dot{\theta} +\left(I \frac{\partial \tilde{A}_{\varphi}}{\partial \varphi} -g \frac{\partial \tilde{A}_{\theta}}{\partial \varphi} \right) \dot{\varphi} + \frac{c}{e}g \dot{P}_{\theta} -\frac{c}{e}I \dot{P}_{\varphi} \right]
\end{eqnarray*}
and the change in parallel momentum as 
\begin{eqnarray*}
\dot{\left(\frac{p_{\para}}{B} \right)} &=& \frac{1}{D_r} \left[ \left(\left(\frac{p_{\para}}{B}g^{\prime} -\frac{e}{c} +\frac{e}{c} \tilde{A}_{\varphi}^{\prime} \right) \frac{\partial \tilde{A}_{\theta}}{\partial \theta} - \left(\frac{p_{\para}}{B} I^{\prime} +\frac{e}{c}q +\frac{e}{c} \tilde{A}_{\theta}^{\prime} \right) \frac{\partial \tilde{A}_{\varphi}}{\partial \theta} \right) \dot{\theta} +\left(1-\tilde{A}_{\varphi}^{\prime} -\frac{c}{e} \frac{p_{\para}}{B} g^{\prime} \right) \dot{P_{\theta}} \right. \\
&& \left. +\left( \left( \frac{p_{\para}}{B} g^{\prime} -\frac{e}{c} +\frac{e}{c} \tilde{A}_{\varphi}^{\prime} \right) \frac{\partial \tilde{A}_{\theta}}{\partial \varphi} - \left(\frac{p_{\para}}{B} I^{\prime} +\frac{e}{c}q +\frac{e}{c} \tilde{A}_{\theta}^{\prime} \right)\frac{\partial \tilde{A}_{\varphi}}{\partial \theta} \right)\dot{\varphi} +\left(q+\tilde{A}_{\theta}^{\prime} +\frac{c}{e} \frac{p_{\para}}{B} I^{\prime} \right) \dot{P}_{\varphi}\right].
\end{eqnarray*}
Setting a constraint to the perturbation of the vector potential
\begin{eqnarray*}
\tilde{\vect{A}} &=& \tilde{\alpha}(\vect{x},t ) \vect{B},
\end{eqnarray*}
one finds that
\begin{eqnarray*}
\tilde{\alpha}(\delta \nabla\psi_p + I\nabla \theta +g \nabla \varphi)&\overset{!}{=}& \tilde{A}_{\psi_p}\nabla\psi_p +\tilde{A}_{\theta} \nabla \theta + \tilde{A}_{\varphi} \nabla \varphi \\
 \Rightarrow \tilde{A}_{\psi_p}& =& \delta \tilde{\alpha}, \\
 \phantom{\Rightarrow} \tilde{A}_{\theta} &=& I \tilde{\alpha}, \\
 \phantom{\Rightarrow} \tilde{A}_{\varphi} &= &g \tilde{\alpha}.
\end{eqnarray*}
Using these expressions in the equations of motion and transferring to SI units
\begin{eqnarray*}
c &\rightarrow& \frac{1}{\sqrt{\mu_0 \epsilon_0}}, \qquad \qquad e \rightarrow \frac{e}{\sqrt{\mu_0 \epsilon_0}}, \\ E &\rightarrow & \sqrt{4 \pi \epsilon_0} E, \qquad \qquad B \rightarrow \sqrt{\frac{4\pi}{\mu_0}} B,
\end{eqnarray*}
we get the following final expressions for the equations of motion:
\begin{eqnarray*}
\dot{\theta}&=& \frac{1}{D_r} \left[ \left(\frac{B^{\prime}}{\gamma} \left(\mu+\frac{B}{m} \left(\frac{p_{\para}}{B}\right)^2 \right)+e\phi^{\prime} \right) \frac{g}{e} + \frac{B^2}{\gamma m} \frac{p_{\para}}{B} \left(1-g^{\prime}\left(\frac{p_{\para}}{eB} + \tilde{\alpha} \right) -\tilde{\alpha}^{\prime} g \right) \right] \\
\dot{\varphi} &=& \frac{1}{D_r} \left[ - \left(\frac{B^{\prime}}{\gamma} \left(\mu+\frac{B}{m} \left(\frac{p_{\para}}{B}\right)^2 \right)+e\phi^{\prime} \right) \frac{I}{e} + \frac{B^2}{\gamma m} \frac{p_{\para}}{B} \left(q+ I^{\prime}\left(\frac{p_{\para}}{eB} + \tilde{\alpha} \right) +\tilde{\alpha}^{\prime} I \right) \right] \\
\dot{\psi}_p &=& \frac{1}{D_r} \left[-\frac{1}{\gamma e} \left(\mu+\frac{B}{m} \left(\frac{p_{\para}}{B}\right)^2 \right)\left(g \frac{\partial B}{\partial \theta}  -I \frac{\partial B}{\partial \varphi} \right) - \left(g \frac{\partial \phi}{\partial \theta} -I \frac{\partial \phi}{\partial \varphi} \right) +\frac{B^2}{\gamma m} \frac{p_{\para}}{B} \left(g \frac{\partial \tilde{\alpha}}{\partial \theta} -I \frac{\partial \tilde{\alpha}}{\partial \varphi} \right) \right] \\
\dot{\left(\frac{p_{\para}}{B}\right)} &=& \frac{1}{D_r} \left[ \left(I \frac{\partial \tilde{\alpha}}{\partial \varphi} - g \frac{\partial \tilde{\alpha}}{\partial \theta}\right)\left(\frac{B^{\prime}}{\gamma} \left(\mu + \frac{B}{m} \left(\frac{p_{\para}}{B} \right)^2 \right) +e\phi^{\prime} \right) \right. \\
&& \left. -\frac{1}{\gamma}\left(\mu+\frac{B}{m} \left(\frac{p_{\para}}{B}\right)^2 \right) \left( \left(1-g^{\prime} \tilde{\alpha} -g \tilde{\alpha}^{\prime} -\frac{p_{\para}}{eB} g^{\prime} \right) \frac{\partial B}{\partial \theta} + \left(q + I^{\prime} \tilde{\alpha} + I \tilde{\alpha}^{\prime} +\frac{p_{\para}}{eB} I^{\prime} \right) \frac{\partial B}{\partial \varphi} \right) \right. \\
&& \left. -e \left(\left(1-g^{\prime}\tilde{\alpha} - g\tilde{\alpha}^{\prime} -\frac{p_{\para}}{eB} g^{\prime} \right)\frac{\partial \phi}{\partial \theta} + \left(q+I^{\prime} \tilde{\alpha} + I\tilde{\alpha}^{\prime} +\frac{p_{\para}}{eB} I^{\prime} \right) \frac{\partial B}{\partial \varphi} \right) \right]
\end{eqnarray*}
where
\begin{eqnarray*}
D_r &=& \frac{p_{\para}}{eB} \left(I^{\prime}g -I g^{\prime} \right) +I +qg +gI^{\prime}\tilde{\alpha} -I g^{\prime} \tilde{\alpha}.
\end{eqnarray*}

\end{appendix}

%\newpage
%\section*{References (shortened for the draft version)}
\addcontentsline{toc}{section}{References}
%\bibliographystyle{unsrt}
% Temporary bst to reduce bibliography length and include url
\bibliographystyle{iaea_papp_natbib}
\bibliography{references}

\end{document}